\documentclass[english,aps,prb,
amsmath,amssymb,preprintnumbers,twocolumn,floatfix,showpacs]{revtex4-1}
\pdfoutput=1
\usepackage{color}
\usepackage[english]{babel}
\usepackage{float}
\usepackage{amsmath}
\usepackage{amssymb}
\usepackage{graphicx}

\makeatletter

 \@ifundefined{textcolor}{}
 {%
   \definecolor{BLACK}{gray}{0}
   \definecolor{WHITE}{gray}{1}
   \definecolor{RED}{rgb}{1,0,0}
   \definecolor{GREEN}{rgb}{0,1,0}
   \definecolor{BLUE}{rgb}{0,0,1}
   \definecolor{CYAN}{cmyk}{1,0,0,0}
   \definecolor{MAGENTA}{cmyk}{0,1,0,0}
   \definecolor{YELLOW}{cmyk}{0,0,1,0}
 }

\usepackage{color}

\usepackage{cancel} 
\usepackage{slashed}

\renewcommand{\emph}[1]{\textit{#1}}

\begin{document}

\title{Dynamics and Hall-edge-state mixing of localized electrons in a two-channel Mach-Zehnder interferometer}
\author{Laura Bellentani}
\email[Corresponding author: ]{laura.bellentani@unimore.it}
\affiliation{Dipartimento di Scienze Fisiche, Informatiche e Matematiche, Universit{\`a} degli Studi 
di Modena e Reggio Emilia, Via Campi 213/A, I-41125 Modena, Italy}
\author{Andrea Beggi}
\affiliation{Dipartimento di Scienze Fisiche, Informatiche e Matematiche, Universit{\`a} degli Studi 
di Modena e Reggio Emilia, Via Campi 213/A, I-41125 Modena, Italy}
\author{Paolo Bordone}
\affiliation{Dipartimento di Scienze Fisiche, Informatiche e Matematiche, Universit{\`a} degli Studi 
di Modena e Reggio Emilia, Via Campi 213/A, I-41125 Modena, Italy}
\affiliation{S3, Istituto Nanoscienze-CNR, Via Campi 213/A, 41125 Modena, Italy}
\author{Andrea Bertoni}
\email[Corresponding author: ]{andrea.bertoni@unimore.it}
\affiliation{S3, Istituto Nanoscienze-CNR, Via Campi 213/A, 41125 Modena, Italy}

\begin{abstract}
We present a numerical study of a multichannel electronic Mach-Zehnder interferometer, based on magnetically-driven non-interacting edge states. The electron path is defined by a full-scale potential landscape on the two-dimensional electron gas at filling factor two, assuming initially only the first Landau level as filled. We tailor the two beam splitters with $50\%$ interchannel mixing and measure Aharonov-Bohm oscillations in the transmission probability of the second channel. We perform time-dependent simulations by solving the electron Schr\"{o}dinger equation through a parallel implementation of the split-step Fourier method and we describe the charge-carrier wave function as a Gaussian wave packet of edge states.
We finally develop a simplified theoretical model to explain the features observed in the transmission probability and propose possible strategies to optimize gate performances.
\end{abstract}


\maketitle

\section{Introduction}
The concrete implementation of quantum information devices is facing a notable development, mainly based on superconducting\cite{Lanting2014_PRX} and single ion\cite{Debnath2016_N} qubits.
Alternative approaches based on electronic states in semiconductor devices seem also to be particularly promising due to their scalability and their potential to be integrated with traditional electronic circuitry. 
However, decoherence represents a major problem for semiconductor devices due to the existence of several scattering sources for electrons in solids, as phonons, impurities, and electron-electron interactions. 
Specifically, for a flying-qubit implementation\cite{Benenti2004, Bertoni2009, Yamamoto2012_NN} of a quantum gate, the onset of environmental interactions would destroy the coherence of the traveling electron wave packet (WP) on very short timescales.

Topologically protected edge states (ESs) are able, in principle, to prevent the loss of coherence of the electron state by embedding it in a subspace that is invariant to small perturbations and is robust against the above scattering mechanisms\cite{Sarma1997}.
For this reason, single electrons in ESs are emergent candidates for the implementation of quantum logic gates\cite{Giovannetti2008_PRB, Beggi2015_JOPCM}.
The most notable example of such states consists of a two-dimensional electron gas (2DEG) subject to an intense transverse magnetic field driving the system into the integer quantum Hall (IQH) regime. In this case, ESs are one-dimensional chiral conductive channels localized at the boundaries between allowed and forbidden regions for the free conduction-band electrons, where the latter can propagate for long distances (larger than $10 \ \mu$m)\cite{Venturelli2011_PRB, Rolleau2008_PRL} without being backscattered, due to the chirality of the channels.
Thus, the IQH regime is an ideal platform for electronic interferometry aimed at quantum information processing, which however requires the realization of semiconductor nanodevices able to manipulate edge channels. Due to the analogy with the corresponding optical systems, this class of systems is often termed \textit{electron quantum optics devices}.
Several examples of the latter have been realized experimentally, such as Mach-Zehnder interferometers (MZI)\cite{Ji2003_N, Deviatov2011_PRB}, Fabry-P\'erot  interferometers\cite{Deviatov2008_PRB, Deviatov2013_LTP, Choi2015_NC}, Hong-Ou-Mandel interferometers\cite{Bocquillon2013_S, Oliver1999_S, Marian2015_JOP, Marguerite2016_PRB, Marguerite2015_NatComm} and Hanbury Brown-Twiss interferometers\cite{Neder2007_N, Bocquillon2012_PRL}, and their application as quantum erasers\cite{Weisz2014_S} or which-path detector\cite{Neder2007_PRL} has been tested. Numerical simulations based on stationary-state\cite{Kreisbeck2010_PRB, Venturelli2011_PRB, Palacios1993_PRB} or time-dependent\cite{Beggi2015_JOPCM, flyingqb_Waintal, Kramer2010_PS} approaches have been essential to understand the experiments, but a time-dependent modeling of a whole IQH device, aiming at the proposal of a quantum gate, is lacking. 

A new promising architecture for a multichannel MZI has been proposed in Ref.~[\onlinecite{Giovannetti2008_PRB}]. Whereas previous MZIs were mainly based on counterpropagating channels\cite{Paradiso2012_PRB} and the typical Corbino geometry\cite{Deviatov2013_LTP}, this device is characterized by a smaller loop area, which reduces the effects of phase-averaging, and, most important, strong scalability, which allows to concatenate in series a number of devices. The system under study operates at filling factor two, contrary to the previous proposal presented in Ref.~[\onlinecite{Beggi2015_JOPCM}], where the device was operating at filling factor one, with a single channel reflected/transmitted by a quantum point contact.
Indeed, numerical studies show that it is possible to realize coherent superposition of edge channels with sharp potential barriers\cite{Palacios1992_PRB}, and that the interchannel mixing coefficient can be arbitrarily tuned by using arrays of top gates\cite{Venturelli2011_PRB, Karmakar2013_JOPCS, Karmakar2015_PRB}. This mechanism has been applied experimentally to spin-resolved edge states\cite{Karmakar2011_PRL, Karmakar2015_PRB}, with an additional in-plane magnetic field to couple the two spin channels.  However, an essential drawback of this idea is the large spatial extension of this beam splitter (BS) and the difficulty in fine-tuning the device operation.
Instead of using spin-resolved ESs, our research focuses on a multichannel MZI where the two non-interacting copropagating channels belong to different Landau levels (LLs), and the formation of the qubit does not require a resonant condition\cite{Venturelli2011_PRB}. As a consequence, in the present device the length in which the two channels need to run on the same edge (i.e. the region at filling factor 2) is limited to a small BS region, as detailed in Appendix~\ref{app:interaction}.
Additionally, we suppose to encode the qubit in a propagating WP of ESs\cite{Beggi2015_JOPCM}, with a Gaussian shape, whose injection protocol for quantum dot pumps has been recently proposed in Ref.~[\onlinecite{Ryu2016_PRL}]. 
This choice corresponds to the experimental situation where a quantum dot pump\cite{Kataoka2016_PRL, Emary2016_PRB}
injects the single-electron WP in the ES of an interferometer, with an energy well above the Fermi energy of the device. 
Though a number of implementations of ES interferometers are based on Lorentzian or exponential WPs\cite{Keeling2006_PRL, Keeling2008_PRL, Dubois2013_N, Feve2007_S}, we found that, in a noninteracting picture, the qualitative results of our simulations are valid also for alternative shapes of the initial wave function. To be more specific, in Appendix~\ref{app:wpshape} we show that our approach applies also to the case of a WP with a Lorentzian distribution in energy\cite{Feve2007_S}.

In order to solve the time-dependent Schr\"odinger equation, we use the split-step Fourier method together with the Trotter-Suzuki factorization of the evolution operator\cite{Kramer2010_PS}, with a parallel implementation of the simulation code\cite{Grasselli2016_PRB}.
Specifically, we study the real-space evolution of the particle state, observing the dynamics of a carrier inside the MZI. We additionally perform support calculations with the Kwant software\cite{Groth2014_NJP}, which solves the scattering problem in a steady-state picture for a single energy component of the WP. 
After optimizing the device and the performance of the quantum gate operation, we measure the transmission probability from the first to the second channel. 
We vary both the length mismatch of the two paths, defined by the width of the mesa $W$, and the orthogonal magnetic field $B$, to observe Aharonov-Bohm (AB) oscillations\cite{Bird1994_PRB, Ji2003_N} in the transmission amplitude. We finally relate the variations of transmission probability in the two outbound channels to the device geometry and compare exact numerical results to a simplified theoretical model based on the scattering matrix formalism.
\begin{figure*}[t]
\begin{centering}
\includegraphics[width=1.\textwidth]{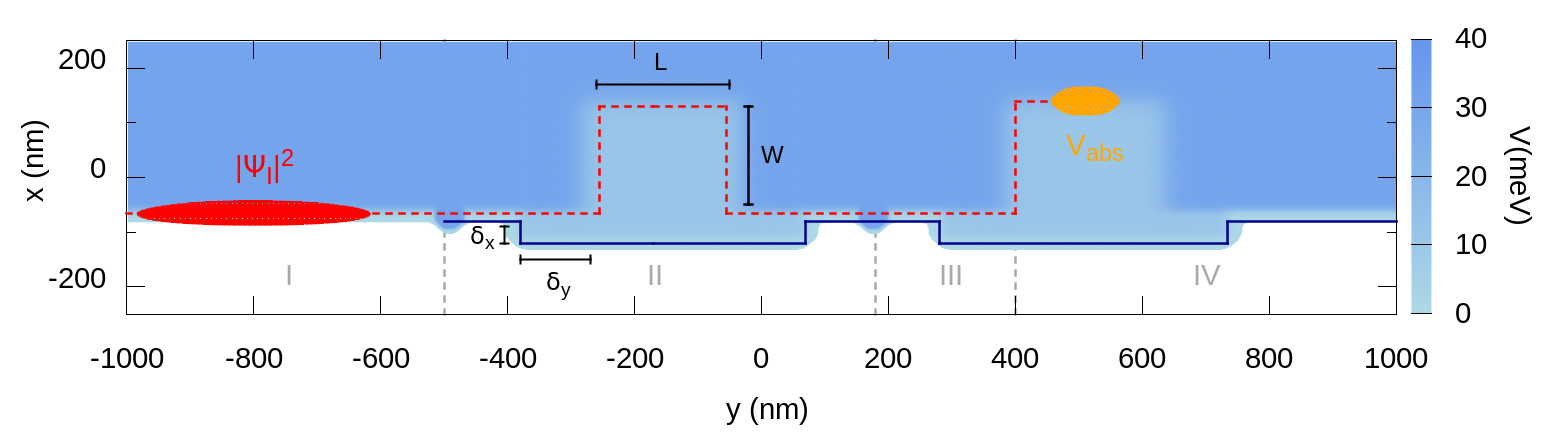}
\par\end{centering}
\caption{Top-view of the two-channel MZI and electron probability density of the initial WP $|\Psi_\text{I}|^2$ in the first edge channel $n=1$. The dashed red ($n=1$) and solid blue ($n=2$) lines describe the intended path of the two edge channels. Region IV contains the measurement apparatus, where the imaginary potential $V_{abs}$ (gold shape at $y=500$~nm) absorbs the first LL. Only the electronic WP in the second channel reaches the right side of the device, where the transmission measurement is performed. The central energy of the WP is about $E_0=20$~meV. }
\label{fig:full_device}
\end{figure*}
\section{Physical system}
In our simulations, a conduction-band electron with charge $-e$ and an effective mass $m^*$  moves in a 2DEG on the $xy$-plane and it is immersed in a uniform magnetic field $\textbf{B}=(0,0,B)$ along the $z$-direction. 
We describe the effect of the magnetic field on the charge carrier in the Landau gauge $\textbf{A}=B(0,x,0)$, that simplifies the definition of the initial state moving along the $y$-direction. 
The potential modulation induced by a polarized metallic gate pattern on the 2DEG is reproduced by the local potential landscape $V(x,y)$ reported in Fig.~\ref{fig:full_device}. The Hamiltonian of a conduction-band electron results to be
\begin{align}\label{Hamilt_xy}
\hat{H}=-\frac{\hbar^2}{2m^*}\frac{\partial^2}{\partial x^2}-\frac{\hbar^2}{2m^*}\frac{\partial^2}{\partial y^2}-i\hbar\frac{eBx}{m^*}\frac{\partial}{\partial y}& \nonumber \\
+\frac{e^2B^2}{2m^*}x^2+V(x,y).&
\end{align}
\par
In order to solve the time-dependent equation of motion, we initialize the electron in a region of the device where $V(x,y)=V(x)$, i.e. where the Hamiltonian shows translational symmetry along the $y$-direction (region I in Fig.~\ref{fig:full_device}), and its eigenstate can be factorized as $\varphi_{n,k}(x)e^{iky}$. 
Following the standard description of the IQH effect, the exponential term $e^{iky}$ describes a delocalized plane wave along $y$, while the function $\varphi_{n,k}(x)$ is the eigenstate of the 1D effective Hamiltonian
\begin{equation}\label{Hamilt_x}
\hat{H}_{1D}^{eff}=-\frac{\hbar^2}{2m^*}\frac{\partial^2}{\partial x^2}+\frac{1}{2}m^*\omega_c^2(x-x_0)^2,
\end{equation}
where $\omega_c=-\frac{eB}{m^*}$ is the cyclotron frequency and
\begin{equation}\label{x0_k0}
x_0(k)=-\frac{\hbar k}{eB}
\end{equation}
is the center of a parabolic confining potential depending on the wave vector $k$ along the $y$-direction. 
The discrete eigenvalues $E_n$ of the effective Hamiltonian (\ref{Hamilt_x}) are the LLs. If the potential $V(x,y)$ is not present, the LL energies do not depend on $k$, while the system eigenfunctions $\varphi_n$ correspond to the Landau states.
On the contrary, the presence of a confining step-like potential $V(x)$ modifies the LL band structure introducing a dispersion on the wave vector, such that $E_n=E_n(k)$. 
In detail, the shape of the eigenstates $\varphi_{n,k}(x)$ changes significantly when the center of the parabolic confinement $x_0$ approaches the nonzero region of $V(x)$. For a fixed $k$, the state $\varphi_{n,k}(x)e^{iky}$ is a current-carrying ES characterized by a net probability flux in the $y$-direction. 
\par
Our time-dependent approach (described later in Section~\ref{sec:IIINumerical-simulations}) requires to go beyond the delocalized description of the wave function. 
In fact, we choose for the initial state a specific value of the $n$ index, and we combine linearly the corresponding ESs on $k$ in order to form a minimum-uncertainty WP. 
From a computational perspective, our choice of a minimum-uncertainty WP as the initial state avoids numerical instabilities and minimizes the real-space spreading of the wave function.
Specifically, the particle is initialized in region I of Fig.~\ref{fig:full_device} in the first edge channel ($n=1$) and is described by
\begin{equation}\label{initial_WP}
\Psi_{\text{I}}(x,y)=\int dk F(k)e^{iky}\varphi_{1,k}(x),
\end{equation}
where the weight function $F(k)$ is the Fourier transform of a Gaussian function along $y$
\begin{equation}\label{weightk}
F(k)=\sqrt[4]{\frac{\sigma^2}{2\pi^3}}e^{-\sigma^2(k-k_0)^2}e^{-iky_0},
\end{equation}
and it entails the wave vector localization around $k_0$.
The Gaussian envelope defines a finite extension around the central coordinate $y_0$, while the localization around $x_0(k)$ is defined by the function $\varphi_{1,k}(x)$.
Unlike our previous work\cite{Beggi2015_JOPCM}, the energy dispersion of the WP includes the energies of the first two edge channels ($n=1,2$) so that, in principle, both can be occupied, even though only the first one is initially filled, as indicated by Eq.~(\ref{initial_WP}). 
\par
The transition between low-potential and high-potential regions in the $x$-direction at fixed $y$ occurs at $x_b$, and we model it by a Fermi-like function. In particular, in the initialization region (region I in Fig.~\ref{fig:full_device}) the potential is assumed to depend only upon $x$, according to the expression
\begin{equation}\label{pot_x}
V(x)=V_b\mathcal{F_{\tau}}(x-x_b),
\end{equation}
where $\mathcal{F_{\tau}}(x)=(\exp(\tau x)+1)^{-1}$ is the Fermi distribution with a smoothness given by the broadening parameter $\tau$, and $V_b$ represents the energy of the forbidden region.
Taking the potential of Eq.~(\ref{pot_x}), the eigenstates $\varphi_{n,k}(x)$ of the effective Hamiltonian are computed numerically. Moving forward along the positive $y$-direction, the potential profile assumes also a dependence on $y$, such that the two edge channels, whose paths are defined by $V(x,y)$, constitute an MZI.
On the border between region I and II in Fig.~\ref{fig:full_device}, the WP impinges on a sharp potential dip, which acts as a BS and redistributes the wave function on the first two available channels $n=1,2$. 
Then, the potential mesa in the middle of region II forces the two channels to follow different paths that accumulate a relative phase.
Proceeding further, between region II and III, a second BS produces the interference between the two parts of the wave function. 
In region III and IV, we introduce an additional mesa and an imaginary potential, respectively, as a measurement apparatus to remove the electron probability from channel $n=1$ alone. As a consequence, the norm of the final wave function represents the total transmission probability of the interferometer from the first to the second channel, $P_{21}^{tot}$.
\section{Numerical simulations
\label{sec:IIINumerical-simulations}}
As previously observed, our time-dependent numerical simulations model the evolution of a localized WP representing the propagating carrier. 
Our method allows to directly observe the dynamics of carrier transport in the time domain and to assess the effects of real-space localization on it. 
This approach does not require the diagonalization of the Hamiltonian of the whole device, which can be a very demanding task for such a large system. 
Indeed, we only perform the diagonalization of the 1D effective Hamiltonian in Eq.~(\ref{Hamilt_x}) with the addition of the confining potential $V(x)$ in region I.
\par
Once the particle is initialized, we solve the time-dependent Schr\"odinger equation by using a parallel implementation of the split-step Fourier method, based on the recursive application of the evolution operator $\hat{U}(\delta t)=e^{-\frac{i}{\hbar}\hat{H}\cdot\delta t}$ to the initial wave function $\Psi_{\text{I}}(x,y;t=0)$:
\begin{equation}\label{eq:psi_init}
\Psi(x,y;t)|_{t=N\delta t}=[\hat{U}(\delta t)]^N\Psi_{\text{I}}(x,y;0) .
\end{equation} 
The kinetic and potential contributions to the Hamiltonian of Eq.~(\ref{Hamilt_xy}) can be split in three parts:
\begin{equation} 
\hat{H}=\hat{T}_1(x,p_y)+\hat{T}_2(p_x)+\hat{V}(x,y),
\end{equation}
where the kinetic terms are defined by
\begin{equation}
\hat{T_1}(x,p_y)=\frac{(p_y-eBx)^2}{2m^*}, \ \ \hat{T_2}(p_x)=\frac{p_x^2}{2m^*}.
\end{equation}
Then, we use Trotter-Suzuki factorization to split the evolution operator, separating the kinetic and potential contributions. 
In order to exploit the diagonal nature of $\hat{T_1}$ and $\hat{T_2}$ on the reciprocal space and of $V(x,y)$ on the real space, we apply alternated Fourier transforms $\textsf{F}_{x(y)}$ and anti-Fourier transforms $\textsf{F}^{-1}_{x(y)}$ along the $x(y)$-direction. 
The evolution operator assumes finally the following form:
\begin{align}
[\hat{U}(\delta t)]^N=&e^{+\frac{i}{\hbar}\delta t\frac{\hat{V}}{2}}[e^{-\frac{i}{\hbar}\delta t\cdot (\hat{V}+\hat{V}_{abs})}\textsf{F}^{-1}_ye^{-\frac{i}{\hbar}\delta t\cdot\hat{T}_1} \nonumber \\ &\textsf{F}_y\textsf{F}_x^{-1}e^{-\frac{i}{\hbar}\delta t\cdot \hat{T}_2}\textsf{F}_x]^Ne^{-\frac{i}{\hbar}\delta t\cdot \frac{\hat{V}}{2}} .
\end{align}
The split-step Fourier method requires a careful choice of the small time step $\delta t$.  In particular $\delta t\ll \frac{\Delta_{x(y)}}{v}$, where $\Delta_{x(y)}$ and $v$ are the real-space grid spacing in the $x(y)$-direction and the group velocity, respectively. Furthermore, to avoid aliasing effects, $\delta t\ll \frac{\hbar}{V}$.
Consistently with the previous requirements, we select an iteration time $\delta t= 10^{-16}$ s. We take the initial state of Eq.~(\ref{eq:psi_init}) with $n=1$, $\sigma=60$~nm and centered at $x_0=-50.9$~nm, $y_0=-800$~nm. We consider GaAs parameters for the hosting material, namely $m^*=0.067m_e$. Furthermore, we use a $2048 \times 4096$ simulation grid. Numerical simulations are performed on a domain including the whole device to study AB oscillations of the transmission amplitude $P_{21}^{tot}$, while a reduced domain is used to study each component of the MZI and optimize gate performances, as reported in the following. 
\subsection{Beam splitter}\label{sec:II-Beam-Splitter}
\begin{figure}[t]
\begin{centering}
\includegraphics[width=1.\columnwidth]{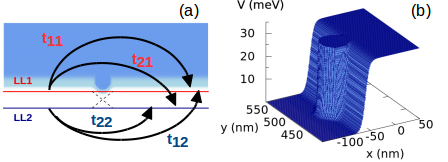}
\par\end{centering}
\caption{(a)Schematic view of the scattering process at the BS. The two horizontal lines represent the two ESs involved in the process (red for the first LL and blue for the second LL), where carriers propagate from left to right. The coefficients $t_{if}$ label the transmission probabilities from the initial state $i$ to the final state $f$. (b) Potential profile modeling the BS.}
\label{fig:scattBS}
\end{figure}
The BS must scatter coherently a particle WP initialized in one of the available channels to fill both LLs and leave the electron in a coherent superposition of the two outgoing channels.
In order to produce the highest visibility of the interferometer, we tune the BS functionality to obtain a $50\%$ mix. Numerical simulations based on delocalized plane-waves\cite{Palacios1992_PRB} show that a coherent edge mixing can be achieved by introducing spatial inhomogeneities on a scale smaller than the magnetic length $l_m$, on the path of the ES. Indeed, an abrupt potential profile scatters elastically an impinging plane wave and redistributes the incoming wave function on the available states (the first two LLs in the present case), with a transmission coefficient $t^{BS}_{f,i}$ from the initial $i=1,2$ to the final $f=1,2$ channels. $t^{BS}_{f,i}$ depends on the energy of the incoming state, on the value of the magnetic field B and on the shape of the local potential. 
\par
Regarding our system, the above mechanism, which is represented in Fig.~\ref{fig:scattBS}(a), is valid for each wave vector component of the particle WP, whose energy distribution is conserved along the whole device. Note that, however, the weight function $F(k)$ depends on the local dispersion of the LLs. In particular, we aim at realizing an edge-channel superposition with equal probabilities, thus requiring a potential profile which ensures a constant transmission probability $|t^{BS}_{12}(E)|^2$=$|t^{BS}_{21}(E)|^2$=0.5 for each energy component $E$ of the initial WP. 
\begin{figure}[b]
\begin{centering}
\includegraphics[width=1.\columnwidth]{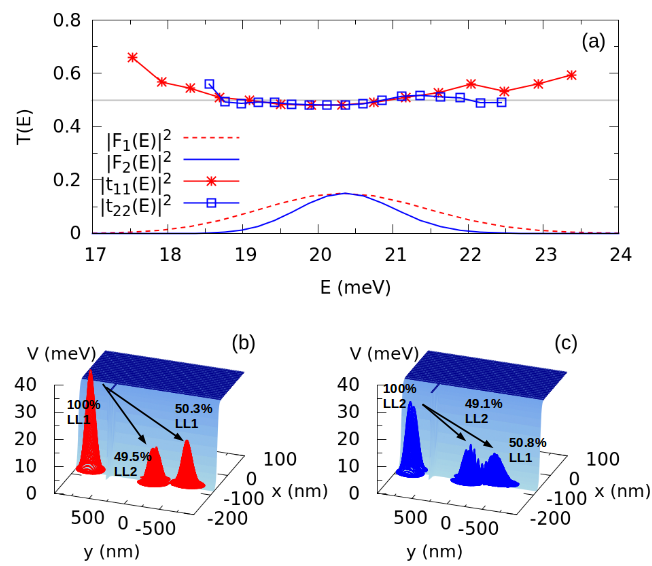}
\par\end{centering}
\caption{(a) Diagonal transmission coefficients $t^{BS}_{ii}$ of the BS as a function of the impinging energy $E$ for $i=1$ (red stars) and $i=2$ (blue squares), calculated with the \textit{wave packet method} and Gaussian envelope function $F(k)$ for a $\Psi$ initialized in $n=1$ (red dashed line) and in $n=2$ (blue solid line). Simulation of the scattering process at the BS for the WP initialized in (b) $n=1$ and (c) $n=2$, where the percentage of the total transmission probabilities is also reported.}
\label{fig:WPmethod}
\end{figure}
We achieve such result by using the potential profile shown in Fig.~\ref{fig:scattBS}(b). Differently from proposals based on spin-resolved ESs\cite{Karmakar2013_JOPCS}, no resonant condition is required. Specifically, our BS consists of a square with the corners smoothed by Fermi profiles:
\begin{align}
V_{BS}(x,y)= V_b \, &\mathcal{F}_{-\tau_{BS}}(x-x_1) \mathcal{F}_{\tau_{BS}}(x-x_2)\nonumber \\ \times &\mathcal{F}_{-\tau_{BS}}(y-y_1) \mathcal{F}_{\tau_{BS}}(y-y_2),
\end{align}
where $\tau_{BS}$ is the smoothing parameter. 
In order to evaluate the energy dependence of $t^{BS}_{if}(E)$, we use the \textit{wave packet method}\cite{Kramer2010_PS}, which is based on a Fourier analysis on the wave function resulting from the scattering at the BS. 
Figure \ref{fig:WPmethod}(a) shows the behavior of $|t^{BS}_{if}(E)|^2$ for $i=f=1,2$, while the interchannel (off-diagonal) coefficients are the complements of the plotted values, due to flux conservation. 
$|t_{11}^{BS}(E)|^2$ and $|t_{22}^{BS}(E)|^2$ are almost constant, with a value close to 0.5, around the central energy of the WP. 
Figure \ref{fig:WPmethod}(a) also reports the energy broadening of the initial WP for a particle initialized in the first (red solid line) and second (blue dashed line) LL. We finally measure the total transmission probabilities simulating the scattering process at the BS with our time-dependent approach. Results are reported in Fig.~\ref{fig:WPmethod}(b) for the WP initialized in the first LL and in Fig.~\ref{fig:WPmethod}(c) for the WP initialized in the second one, at B= 5 T.
A small scattering to the third LL, whose energy is slightly reached by the energy broadening of our initial WP, explains the discrepancy between the sum of the two scattered intensities and unity. 
\par
\begin{figure}[t]
\begin{centering}
\includegraphics[width=1.\columnwidth]{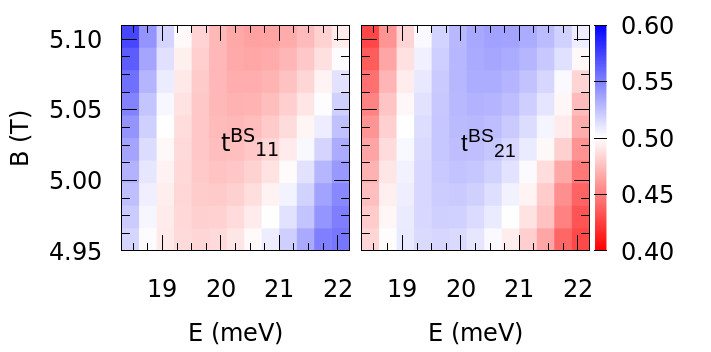}
\par\end{centering}
\caption{Energy and magnetic dependence of the transmission coefficients $t_{11}$ (left panel) and $t_{21}$ (right panel) of the BS in Fig.~\ref{fig:scattBS}(b), obtained with Kwant software.}
\label{fig:scattmap}
\end{figure}

Finally, we perform support calculations with Kwant software\cite{Groth2014_NJP}, simulating delocalized ESs impinging on the BS. The scattering matrix method is used to calculate the maps of Fig.~\ref{fig:scattmap}, where the probabilities $|t_{f,i}^{BS}|^2$ are reported also as a function of the magnetic field $B$. The latter results confirm the transmission probabilities of Fig.~\ref{fig:WPmethod}(a) obtained with the time-dependent method and shows how $B$ tailors the transmission coefficients. 
\subsection{The MZI\label{sec:III-The-MZI}}
\begin{figure}[b]
\begin{centering}
\includegraphics[width=1.\columnwidth]{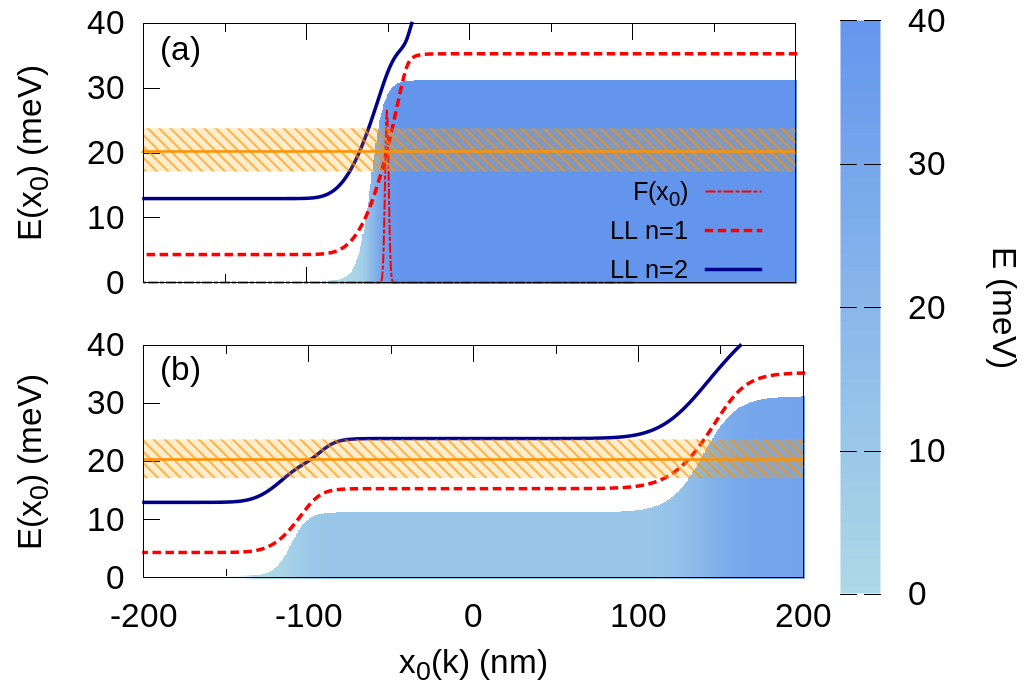}
\par\end{centering}
\caption{Computed real-space band structure of the first two LLs in the two-channel MZI. (a) Band structure of the device at $y=-800$~nm (region I) compared to the potential profile of the confining wall. The orange striped interval defines the energy broadening of the initial WP. Only the first LL is filled with $F(x_0)$, as illustrated by the dotted Gaussian curve around $x_0=-50$~nm. (b) Band structure of the device at $y=-200$~nm (region II) and potential profile of the mesa structure. The energy broadening fills the two edge channels at the two extremes of the potential, inducing two different paths for $n=1$ (red dashed line) and $n=2$ (blue solid line), localized around $x_0=120$~nm and $x_0=-100$~nm, respectively.}
\label{fig:dev_bands}
\end{figure}
Once the coherent superposition is realized, the MZI requires that the two channels accumulate a relative phase. This can be induced by a mismatch of the path lengths or by a net flux of the magnetic field through the \textit{loop area}, which is the area enclosed by the paths of the two channels. To separate the channels, we introduce an area where the potential $V(x,y)$, which mimics the landscape of polarized top gates, has an energy value $V_s$ in between the first and the second LL. In order to avoid an unwanted mix of the two channels and to better model a real device, we create a smooth transition between the two regions by means of the following function:
\begin{equation}
V_G(x,y)=V_s\mathcal{F}_{\tau_s}(x-x_s)+V_b\mathcal{F}_{\tau_b}(x-x_b).
\end{equation}
The smoothness of the local curvature $\tau_s$ must ensure an adiabatic separation of the two edge channels\cite{Venturelli2011_PRB}, with a negligible mixing among them. 
This creates a region (lighter blue in Fig.~\ref{fig:full_device}) where the filling factor is one, in contrast to the bulk filling factor of two.
\par
From a different perspective, in order to split the two channels, we exploit the relation between the real coordinate $x_0$, defining the center of the WP along x, and the momentum $k$ of the traveling particle along y, as given by Eq.~(\ref{x0_k0}).
Indeed, the band structures of the LLs are strictly related to the shape of the potential profile, as shown in Fig.~\ref{fig:dev_bands}(a) for the region I and Fig.~\ref{fig:dev_bands}(b) for a section of the mesa structure in region II. In detail, in Fig.~\ref{fig:dev_bands}(b) it is clear that the potential step pushes upwards the local band structure, and the two LLs are then filled at different $k$. The elasticity of the scattering process at the mesa ensures that the first LL is filled on top of the step potential, while the second LL intersects the energy window at its bottom. The channels are therefore forced to follow a different path, whose length can be tuned by changing the width of the mesa $W$.
The simultaneous recollection of the WPs at the second BS, which is needed to observe the interference, could be prevented by the different group velocity of the WPs in the two edge channels. Indeed the group velocity of the first channel is larger than the group velocity of the second one due to the different band structures of the two LLs. Therefore, we introduce a sort of indentation in the forbidden region on the mesa (region II in Fig.~\ref{fig:full_device}), in order to increase the length of the channel $n=1$ and compensate this effect. Additionally, we smooth the local confining potential inside the indentation, in order to reduce the group velocity of the WP in $n=1$.  
\par
\begin{figure*}[ht]
\begin{centering}
\includegraphics[width=1.\textwidth]{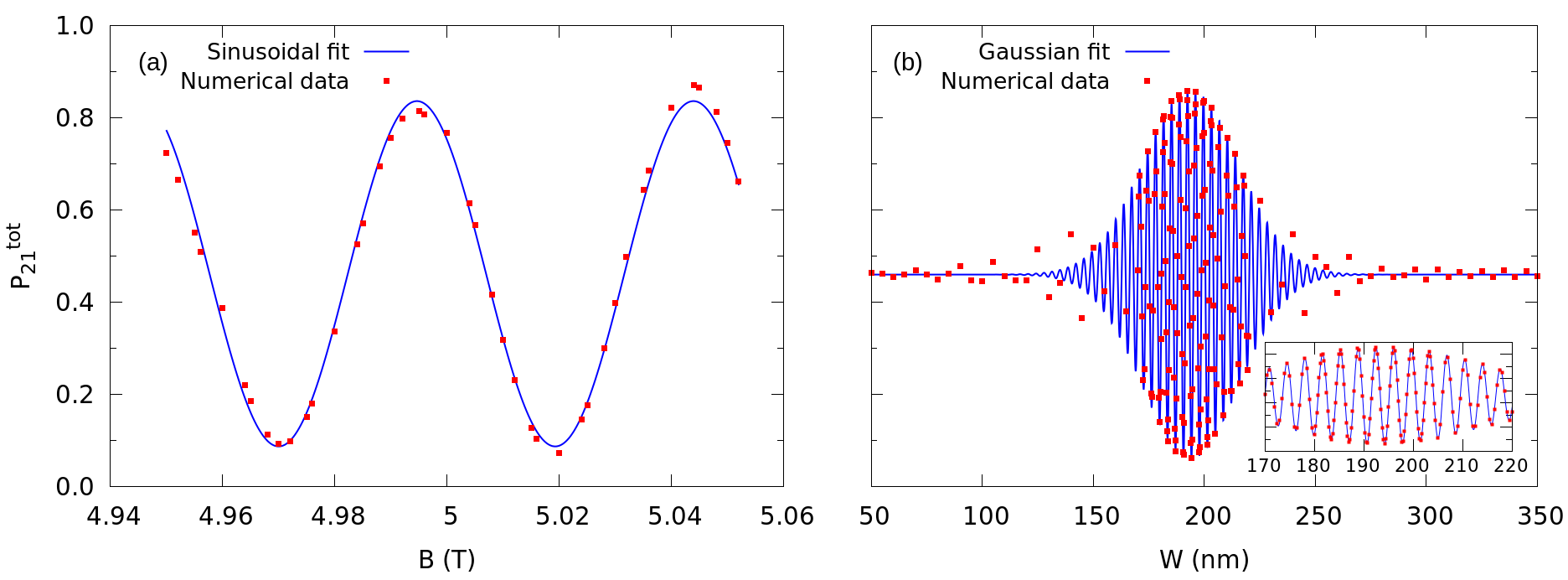}
\par\end{centering}
\caption{AB oscillations of the transmission amplitude at the end of the device. (a) The plot shows the behavior of $P^{tot}_{21}$ as a function of the magnetic field for the numerical simulation (dots) and its sinusoidal fit (blue line) based on the theoretical model of Sec.~\ref{sec:theoreticalmodel}. A fixed path mismatch for the two channels, $W=200$~nm, is considered. (b) AB oscillations as a function of the width of the mesa $W$ at $B=5$~T. Numerical data (dots) and the Gaussian fit (solid line) based on the theoretical model show a good agreement only in the central region (inset). }
\label{fig:AB}
\end{figure*}

Finally, the regions III and IV of the device correspond to the measurement apparatus. After the interference, the two channels are separated by an additional mesa in region III. 
In order to remove from the device the part of the wave function occupying the first LL after the MZI, we introduce the absorbing imaginary potential 
\begin{equation}
V_{abs}(x,y)=iV_{abs}^0\frac{\mathcal{F}_{\tau}(x_a-x)\mathcal{F}_{\tau}(x_b-x)}{\cosh^2((\frac{y-y_c}{d})^2)},
\end{equation} 
where $y_c$ defines its center, $d$ its length, $V^0_{abs}$ its maximum, and $x_a$, $x_b$ define its spatial extension in the $x$-direction. This potential is represented by the gold shape in region IV of Fig.~\ref{fig:full_device} at $y=500$~nm and models a metallic absorbing lead on the path of the first LL. Consequently, the surviving part of the final
wave function gives the probability for the electron to be transmitted in the second LL by the interference process taking place inside the device.
Using the split-step Fourier method, we finally simulate the interference for different values of the orthogonal magnetic field $B$ at $W=200$~nm, and for different widths of mesa $W$ at $B=5$~T, modifying the magnetic and the dynamic phase respectively. Numerical simulations have been performed considering $V_b=0.031$~eV, $\tau_b=0.25$ $\text{nm}^{-1}$ for the confining potential, $|x_1-x_2|=|y_1-y_2|=20$~nm and $\tau_{BS}=0.5$ $\text{nm}^{-1}$ at the BS, $V_s=0.011$~eV, $\tau_s=0.2$ $\text{nm}^{-1}$ for the mesa structure and $V_{abs}^0=-100$~eV, $d=30$~nm for the absorbing potential. The numerical results are reported in Fig.~\ref{fig:AB}. We observe AB oscillations in the transmission amplitude with an high visibility, defined as
\begin{equation}
\nu_{MZI}=\frac{I_{max}-I_{min}}{I_{max}+I_{min}}=\frac{T_{max}-<T>}{<T>},
\end{equation}
thanks to the optimization of the scattering process at the BS. 
Before discussing the results, in the following section we propose a simplified theoretical model whose predictions will help in understanding the outcomes of the exact time-dependent approach.
\section{Theoretical model}\label{sec:theoreticalmodel}
Here we present a theoretical model based on the description of edge channels as strictly one-dimensional systems, using the scattering matrix formalism.  
An ES of the $n^{th}$ LL is represented by a plane wave along $y$, $|k,n\rangle$, with the energy dispersion of that LL, $k(E,n)$.
In order to introduce particle localization on the $y$-direction, our initial wave function is computed by combining different ESs of the $n=1$ level, with the Gaussian weight $F(k)$ of Eq.~(\ref{weightk}):
\begin{align}
|\Psi_{\text{I}}\rangle&=\int dk  F(k) |k,n=1\rangle \nonumber \\
        &=\int dE F(k(E,1)) \left[\frac{dk}{dE}\right]_{n=1}|E,1\rangle,
\end{align} 
where $|E,n\rangle$ denotes $|k(E,n),n\rangle$ for brevity, and $|\Psi_{\text{I}}\rangle$ ($|\Psi_{\text{III}}\rangle$) is the one-dimensional wave function in region I (III). 
We assume a bulk filling factor of two, so that $n$ can be either $1$ or $2$ and represents a pseudo-spin degree of freedom.
The WP in region III can be related to the initial one by describing the scattering process through the application of three operators:
\begin{equation}
|\Psi_\text{III}\rangle=\hat{B}\hat{\Phi}\hat{B}|\Psi_\text{I}\rangle,
\end{equation}
where $\hat{B}$ describes the effect of a BS, and $\hat{\Phi}$ the relative phase accumulated by the two channels in the mesa region. Here, differently from the full numerical simulation of the previous sections, the energy-dependence of $\hat{B}$ and $\hat{\Phi}$ is neglected for simplicity. 
Finally, since the absorbing potential in region IV collects the contribution of the first LL, only $n=2$ survives, and the total transmission probability at the end of the device is defined by the following equation:
\begin{align}\label{eq:p21}
P_{21}^{tot}=&\int dE \left|\langle E,2|\Psi_\text{III}\rangle\right|^2= \nonumber \\
=&\int dE \left|F(k(E,1))\left[\frac{dk}{dE}\right]_{n=1}\right|^2|\langle E,2|\hat{B}\hat{\Phi}\hat{B}|E,1\rangle|^2.
\end{align}
In order to solve Eq.~(\ref{eq:p21}), we consider the general 2x2 matrix form of operators $\hat{B}$ and $\hat{\Phi}$ on the pseudo-spin basis:
\begin{equation}
\hat{B}=
\begin{pmatrix}
b_{11}&b_{12}\\
b_{21}&b_{22}
\end{pmatrix},
\ \ \
\hat{\Phi}=\left(\begin{array}{cc}
e^{i\varphi_1} & 0\\
0 & e^{i\varphi_2}
\end{array}\right).
\end{equation}
The phase $\varphi_i$ ($i=1,2$) includes the contributions of the magnetic ($\phi_i$) and the dynamical ($\xi_i$) phases 
\begin{equation}\label{eq:phases}
\varphi_i=\frac{1}{\hbar}\int_i(p-qA)\cdot ds=\xi_i+\phi_i,
\end{equation} 
where the integration is performed along the path of the edge channel $i$ on the mesa.
The transmission coefficients $b_{ij}$ are related by the probability flux conservation:
\begin{eqnarray}\label{pflux}
|b_{ii}|^2+|b_{ij}|^2&=&1, \\
|b_{ij}|^2&=&|b_{ji}|^2, \\
|b_{ii}|^2&=&|b_{jj}|^2.
\end{eqnarray}  
As in the previous sections, we tune the BS to $50\%$ transmission, so that all the coefficients $|b_{ij}|^2=0.5$. Therefore, $b_{11}$ and $b_{22}$ only differ by a phase factor $\varphi$, such that $b_{22}=b_{11}e^{i\varphi}$.
The transmission probability from channel 1 to channel 2 for a given energy $E$ is
\begin{equation}
|\langle E,2|\hat{B}\hat{\Phi}\hat{B}|E,1\rangle|^2=2|b_{21}b_{11}|^2[1+\cos(\Phi)].
\end{equation}
In order to define a gauge-independent dynamical phase, we consider a quasi-linear dispersion of the two LLs around the central energy of the WP $E_0$, and we rewrite $p$ in terms of the constant energy $E$ and of the group velocity $v_i^\text{II}$ in region II:
\begin{equation}
\xi_i=\frac{1}{\hbar}\int_i\frac{E-E_0}{v_{i}^\text{II}}ds=\frac{E-E_0}{\hbar v_{i}^\text{II}}\Delta S_i,
\end{equation}
where $\Delta S_i$ is the length of the path of channel $i$ in the mesa region. Note that, while considering a linear dispersion is appropriate for the second LL, it represents an approximation for the first one. Such assumption is the main source of discrepancy between our exact numerical results and the present theoretical model.  
Using the Stokes theorem for the magnetic contribution $\phi_i$ of Eq.~(\ref{eq:phases}), we can rewrite the total phase as
\begin{equation}\label{def_phi}
\Phi=\varphi+\frac{E-E_0}{\hbar}\left( \frac{\Delta S_2}{v_{2}^\text{II}} - \frac{\Delta S_1}{v_{1}^\text{II}}\right)+\frac{eB\mathcal{A}}{\hbar},
\end{equation}
with $\mathcal{A}$ the area enclosed by the paths of the two channels, which is tuned by changing the width $W$ of the mesa along the $x$-direction.
Performing the integration over the energy in Eq.~(\ref{eq:p21}), the total transmission probability from channel 1 to channel 2 is
\begin{equation}
P_{21}^{tot}=\frac{1}{2}\left(1+\exp\left(-\frac{(\frac{\Delta S_2}{v_{2}^\text{II}}-\frac{\Delta S_1}{v_{1}^\text{II}})^2}{8\sigma^2/(v_{1}^\text{I})^2}\right)\cos(\Phi')\right),
\end{equation}
where the argument of the cosine $\Phi'=\frac{eB\mathcal{A}}{\hbar}+\varphi$
exposes the dependence of $P_{21}^{tot}$ on the magnetic field $B$ and on the width $W$ of the mesa.
Indeed, according to the geometry of the step potential in Fig.~\ref{fig:full_device}, the mesa has an area $\mathcal{A}=W\cdot L + (2\delta_y+L)\cdot\delta_x$, such that the two following definitions of $\Phi'$ hold:
\begin{align}
\Phi'&=\left( \frac{eBL}{\hbar} \right) W+\Phi_0=k_W W +\Phi_0 \\
     &=\left( \frac{e\mathcal{A}}{\hbar} \right) B+\Phi_1=k_B B +\Phi_1,  
\end{align}
where $\Phi_0=\varphi+\frac{eB}{\hbar}(2\delta_y+L)\delta_x$ and $\Phi_1=\varphi$.
Besides, according to Fig.~\ref{fig:full_device}, the paths of the two channels are equivalent to $\Delta S_1=2\delta_y+2W+L$ and $\Delta S_2=2\delta_x+2\delta_y+L$, and using an effective standard deviation $\Sigma=\sigma v_1^\text{II}/v_1^\text{I}$, the total transmission probability is
\begin{equation}\label{eq:totp21}
P_{21}^{tot}=\frac{1}{2}\left(1+\exp\left(-\frac{(W-W_0)^2}{2\Sigma^2}\right)\cos(\Phi')\right),
\end{equation} 
with $W_0$ containing the geometrical correction to the paths of the two edge channels.
\section{Discussion}
\begin{table*}[htbp]
  \centering
  \begin{tabular}{p{0.16\linewidth} p{0.16\linewidth} p{0.16\linewidth} | p{0.16\linewidth} p{0.16\linewidth} p{0.16\linewidth}}
  \hline
\multicolumn{3}{c|}{\textbf{$P_{21}^{tot}(B)$}} & \multicolumn{3}{|c}{\textbf{$P_{21}^{tot}(W)$}}\\
\hline
expression & Numerical fit & Theoretical model & 
expression & Numerical fit & Theoretical model\\
    \hline
    $A_B$ & 0.462$\pm$0.004 & 0.5 & $A_W$ & 0.460$\pm$0.002 & 0.5\\
    $A_B^*$ & 0.374$\pm$0.005 & 0.5 & $A_W^*$ & 0.400$\pm$0.004 & 0.5\\
    $k_B$ ($\text{T}^{-1}$) & 127.4$\pm$0.4 & 110 & $k_W$ ($\text{nm}^{-1}$) & 1.750 & 1.9\\
    $B_1$ (T) & 4.982$\pm$0.001 & - & $L_1$ (nm) & 192.5 & -\\
    & & & $\Sigma$ (nm) & 21.7$\pm$5 & 18.3\\
     & & & $L_0$ & 193.8$\pm$0.3 & -\\ 
  \end{tabular}
   \caption{Comparison between fitting parameters for the results of exact numerical simulations and the corresponding parameters of the theoretical model of Sec. \ref{sec:theoreticalmodel}. The two cases of Fig.~\ref{fig:AB} are considered, namely with a variable magnetic field (left column) or mesa width (right column).}\label{tab:1}
\end{table*}
The AB oscillations simulated numerically are compared to the transmission probability $P_{21}^{tot}$ of Eq.~(\ref{eq:totp21}) predicted by our theoretical model. In detail, the numerical data are fit by the function
\begin{equation}\label{eq:p21b}
P_{21}^{tot}(B)=A_B+A_B^*e^{-\frac{(W-W_0)^2}{2\Sigma^2}}\cos(k_B(B-B_1))
\end{equation}
for a variation of the magnetic field $B$ [Fig.~\ref{fig:AB}(a)], and by the function
\begin{equation}\label{eq:p21w}
P_{21}^{tot}(W)=A_W+A_W^*e^{-\frac{(W-W_0)^2}{2\Sigma^2}}\cos(k_W(W-W_1))
\end{equation}
for a variation of the width $W$ of the mesa region [Fig.~\ref{fig:AB}(b)]. The comparison between numerical and theoretical parameters is presented in Tab.~\ref{tab:1}.
\par
Regarding the magnetically-driven AB oscillations in Fig.~\ref{fig:AB}(a), we observe that the shape of the interference curve does not describe a perfect sinusoid, but the amplitude slightly increases with the magnetic field. Indeed, an increase of $B$ enhances the spacing between the two LLs, reducing the unwanted interchannel mixing at the step potential, therefore increasing the oscillation visibility. Additionally, our theoretical model neglects the dependence of the transmission coefficients $b_{if}$ on $B$. Fig.~\ref{fig:scattmap} shows indeed that an increase of the magnetic field increases the scattering from the first to the second channel at the BS, affecting the values of $A_B$ and $A^*_B$ in Eq.~(\ref{eq:p21b}). The underestimation of the pseudo periodicity $k_B$ is induced by the approximation of the loop area $\mathcal{A}$, which doesn't take into account the small difference in the $x$ position of the two channels also in the regions with filling factor two.
\par
The amplitude of $P_{21}^{tot}(W)$ in Fig.~\ref{fig:AB}(b) has a damping induced by the relative dynamical phase together with the finite dimension of the wave function. Indeed, when the width $W$ of the mesa is large enough, the two WPs do not overlap anymore and the interference is quenched\cite{Haack2011_PRB}. Such damping was observed also in the single-channel MZI\cite{Beggi2015_JOPCM}, but in the present device $\Sigma$ is reduced with respect to the standard deviation of the initial WP, $\sigma$. In fact, in this two-channel MZI, the smoother slope of the indentation in region II reduces the group velocity $v^\text{II}_1$ above the mesa with respect to $v^\text{I}_1$. This can be interpreted as an effective dilatation of the width $W$ in Eq.~(\ref{eq:totp21}), determining a larger phase difference.
Moreover, as shown in the inset of Fig.~\ref{fig:AB}(b), the Gaussian fit describes properly the oscillation amplitude of $P_{21}^{tot}$ only in the central region, while on the two sides the AB oscillations are larger than the predicted ones.
\par
We measure a visibility $\nu=0.87$ at $W=W_0$ in place of 1, as a consequence of the energy dependence of the phase factors and of the scattering processes inside the device. In particular, in addition to neglecting the energy dependence of the transmission probability at the BS, our theoretical model does not take into account the unwanted interchannel mixing induced by the step encountered by the second LL when entering the mesa region. The mix is actually non zero, and it depends on the energy of the impinging WP. We expect that the high-energy components of the wave function in the second LL [top of the orange striped zone in Fig.~\ref{fig:dev_bands}(b)] are transferred more easily to the states of the first LL with the same energy and a higher group velocity, leaving sooner the scattering region. 
\par
%
%

In summary, in this paper we have investigated the transport properties of a Gaussian electronic WP in a two-channel MZI in the IQH regime. Our numerical modeling of the device required the definition of a proper potential landscape $V(x,y)$ to ensure a high visibility of the transmission amplitude. A specific design of the BS has been used to separate the impinging state into a $50\%$ coherent superposition of the two available channels. However, we found that the proper function of the BS is preserved when different shapes of the mixing potential are used, as we show in the Appendix~\ref{app:wpshape}. We observed AB oscillations, relating the features of transmission-probability amplitude to particle localization, which is inherent in our time-dependent solution. Finally, our numerical results are clarified by a simplified theoretical model based on the scattering matrix formalism and a one-dimensional model for chiral transport in edge states. 
We emphasize that this implementation of an MZI solves the scalability problem\cite{Giovannetti2008_PRB} of the single-channel MZI we studied in Ref.~[\onlinecite{Beggi2015_JOPCM}], thus potentially enabling its concatenation in series and its integration into sophisticated quantum computing architectures. The possibility to concatenate two or more MZI in series, exploiting as an input the two possible outputs of a previous interferometer, is essential for the implementation of two-qubit interferometers, as the Hanbury-Brown-Twiss one\cite{Neder2007_N}, where interfering identical Gaussian WPs could be, in principle, generated from nonidentical sources\cite{Ryu2016_PRL}. In addition, the present device shows a larger visibility with respect to our previous single-channel interferometer\cite{Beggi2015_JOPCM}, mainly due to the weak energy selectivity of the present BS compared to the quantum point contact. Moreover, our BS does not require the resonant condition of the spin-resolved multichannel MZI proposed in Ref.~[\onlinecite{Karmakar2015_PRB}], thus reducing the interchannel interaction induced by the spatial extension of the top gate array (Appendix~\ref{app:interaction}).  

\section*{ACKNOWLEDGMENTS}
We thank G Sesti, I Siloi and E Piccinini for fruitful discussions. 
We acknowledge CINECA for HPC computing resources and support under the ISCRA initiative (IsC48 MINTERES). PB and AB thank Gruppo Nazionale per la Fisica Matematica (GNFM-INdAM).
\appendix

\section{Role of interchannel interactions}\label{app:interaction}

\begin{figure*}\centering
\includegraphics[width=0.95\textwidth]{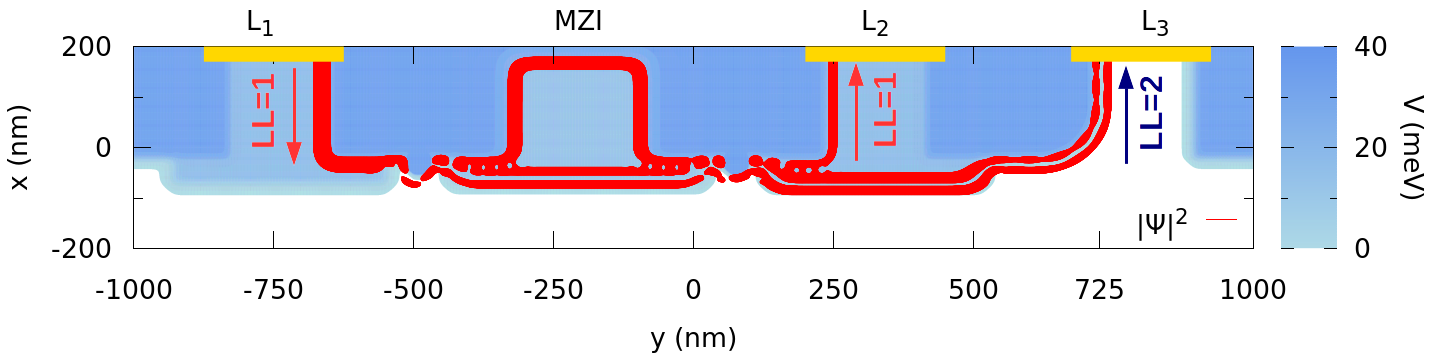}\caption{A time-independent simulation performed with Kwant software\cite{Groth2014_NJP}. Top-view of the MZI with injection ($L_1$) and absorption ($L_2$ and $L_3$) leads. $L_1$ and $L_2$ have unitary bulk filling factor, while $L_3$ has filling factor two. The red profile is the electron density probability for an energy $E=20.4$~meV, injected in the first edge channel from lead $L_1$, and $B=5$~T.}\label{fig:kwantdevice}
\end{figure*}
A large number of experiments\cite{Marguerite2016_PRB, Neder2006_PRL, Huyn2012_PRL} show that at filling factor two, the occurrence of interchannel interactions affects the coherence of the traveling electron. These interactions lead to charge fractionalization, whose effects were exposed in experiments on traditional Mach-Zehnder interferometers\cite{Huyn2012_PRL} and then rationalized by Ref.~[\onlinecite{Helzel2015_PRB}]. In the latter studies, the two available ESs copropagate for very large distances, so that the injected electrons interact with the Fermi sea of the other channel. On the contrary, in the geometry of the MZI proposed in Ref.~[\onlinecite{Giovannetti2008_PRB}], the separation of the two edge channels by a potential mesa quenches interchannel interactions, that arise only at the BS\cite{Chirolli2013_PRL}. However, the significant spatial extension of the BS devised by Karmakar et al. in Ref.~[\onlinecite{Karmakar2015_PRB}] introduces non-negligible interchannel interactions affecting the visibility of the AB oscillations. 

In our implementation of the MZI we propose a single potential dip as a BS with a smaller spatial extension (about $20$~nm). We also remark that, in order to reduce the length of copropagation, the injection and collection of the two channels can be performed using top gates, as in Ref.~[\onlinecite{Karmakar2015_PRB}]. Fig.~\ref{fig:kwantdevice} shows the result of a numerical simulation performed with Kwant software, where only the central energy of the WP is injected. Here the lead $L_1$ has a unitary bulk filling factor and injects the electron in the first edge channel, while lead $L_2$ and lead $L_3$ adsorb the first and second edge channels, respectively. 
Following Ref.~[\onlinecite{Ferraro2014_PRL}], the length over which fractionalization arises is connected to the emission time and group velocity of the WP. We stress that in our computations we chose Gaussian WP with an energy broadening much larger than typical experimental one\cite{Marguerite2016_PRB, Bocquillon2013_S}. This ensures that the selectivity of the BS is still adequate - and actually even better optimized - for larger WPs. For example, we performed numerical simulations for a Gaussian WP with $\sigma=100$~nm, whose energy broadening (full width at half maximum) is $\Gamma=1$~meV and velocity is $v_g=100 \ \text{nm}/\text{ps}$. If we assume an emission time $\tau_e=\hbar/\Gamma$, the length of charge fractionalization\cite{Ferraro2014_PRL} results to be $L_{frac}=v_g \cdot \tau_e \approx 0.07 \ \mu$m, which is larger than the length of the BS region of our device. We expect that for larger emission times, as the ones typically exploited for experimental implementations of single electron sources\cite{Marguerite2016_PRB, Bocquillon2013_S}, $L_{frac}$ is wider (typically $3 \ \mu$m\cite{Ferraro2014_PRL}), while the size of our BS is even more optimized to produce a $50\%$ interchannel mixing. This implies that a proper shape of top gates as in Fig.~\ref{fig:kwantdevice}, together with the use of our type of  BS, could quench significatively the effect of interchannel interaction and avoid, or at least strongly reduce, this source of decoherence, without affecting the performances of the device. 

\section{Alternative shapes for the WP and the BS}\label{app:wpshape}
\begin{figure}\centering
\includegraphics[width=1\columnwidth]{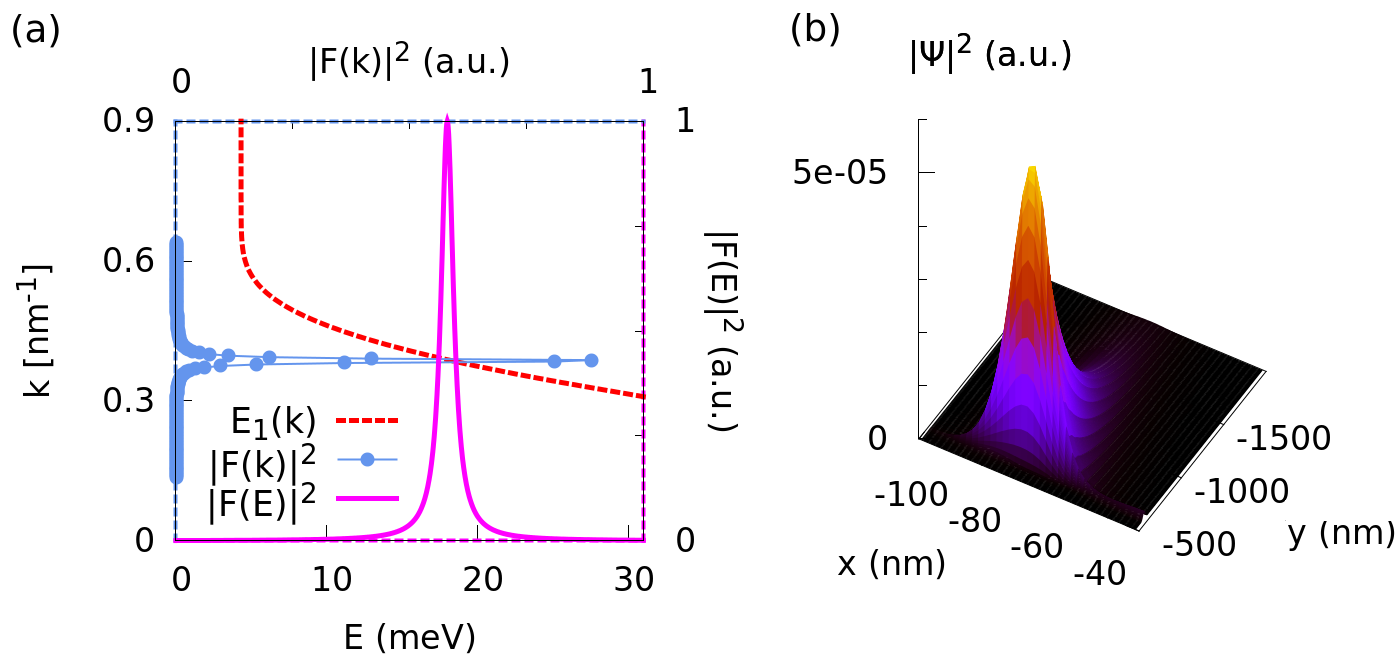}\caption{Lorentzian WP at $t=0$~ps. (a) Energy distribution (magenta), $k-$space distribution (light blue), and dispersion  curve of the first LL $E_1(k)$ (red dashed line); (b) probability density in the real $xy$-space.}\label{fig:lorwp}
\end{figure}
\begin{figure}\centering
\includegraphics[width=1\columnwidth]{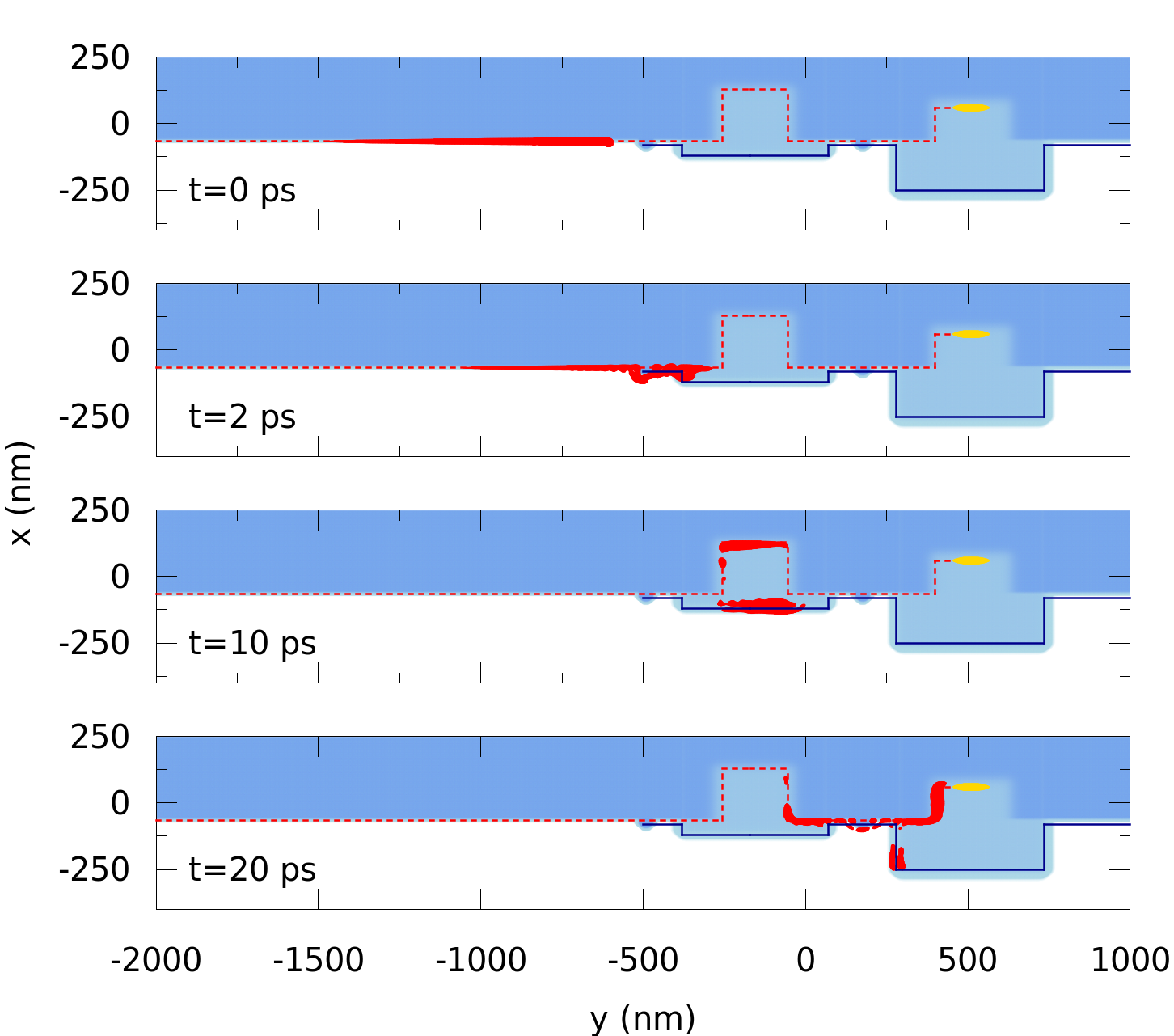}\caption{Snapshot in time of the Lorentzian WP with $\Gamma=0.001$~eV (red curve) in the potential landscape defining the MZI (blue map). Note the asymmetric shape of the initial WP and the large spread of the wave function during its propagation.}\label{fig:lorev}
\end{figure}
The functioning of our MZI does not depend on the specific shape of the WP. The choice of a Gaussian weight function is motivated by the higher control of its time evolution with respect to alternative shapes. For the sake of completeness, here we show the evolution of a WP with a Lorentzian distribution in energy, in order to mimic the emission of electrons by a mesoscopic capacitor\cite{Feve2007_S}. Fig.~\ref{fig:lorwp} shows the initial broadening of the WP in energy and real space: the two long tails of the Lorentzian distribution produce a small filling of the states with no velocity and collect a very large number of wave vectors, thus inducing a larger spread of the WP during its evolution. Additionally, due to its very small energy peak, the wave function in real space has a long tail, that required to double the dimensionality of the initialization region. Fig.~\ref{fig:lorev} shows its evolution at different time steps: the initial beam in the first edge channel ($t=0$~ps) is split in a coherent superposition of the two channels by the first BS ($t=2$~ps), than the mesa structure separates the component with different $n$ ($t=10$~ps), and finally the WPs are recollected at the second BS ($t=20$~ps) to realize the interference. Numerical results confirm that our device is still fully operational in this case.

\begin{figure}\centering
\includegraphics[width=0.75\columnwidth]{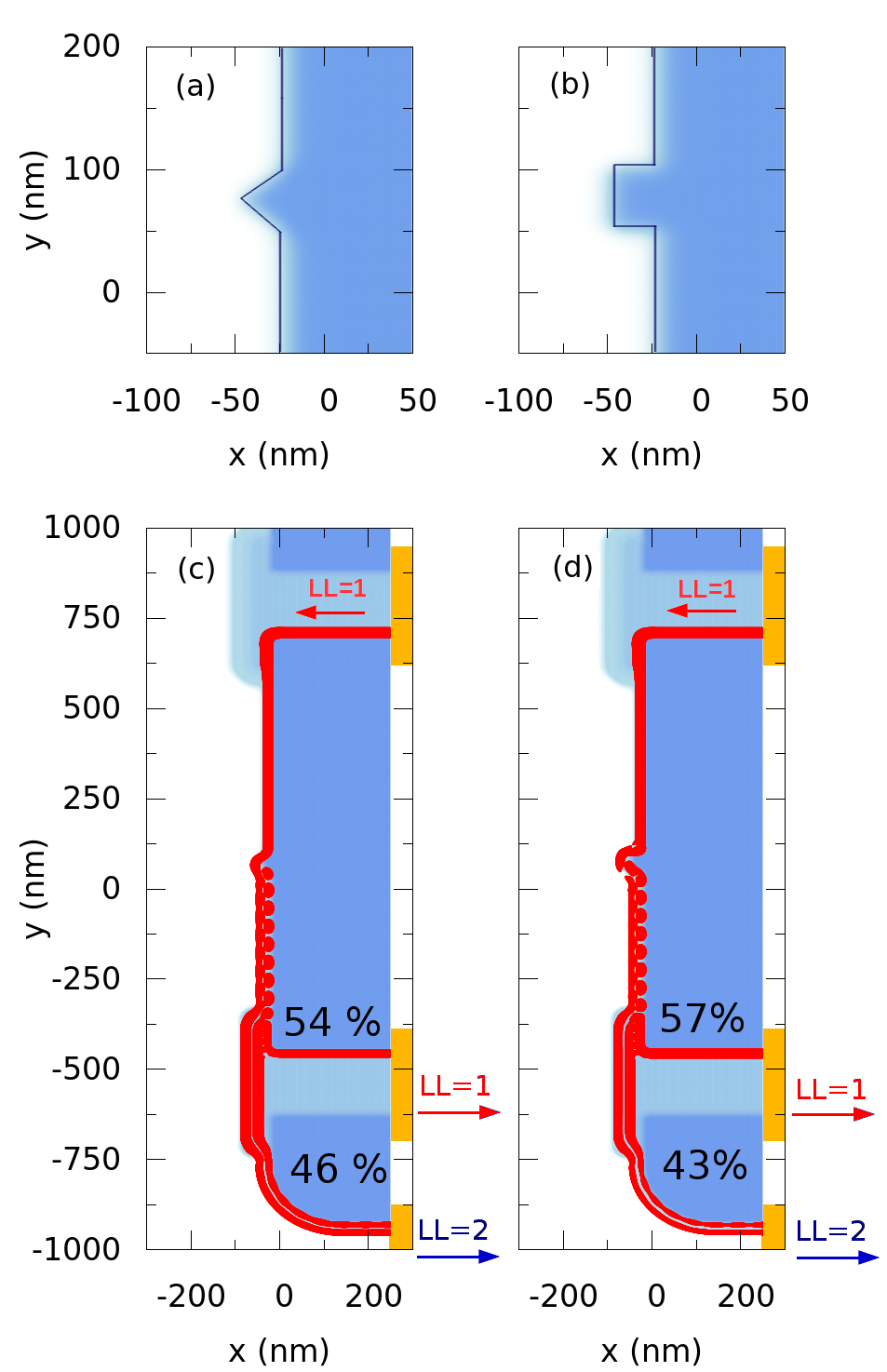}\caption{Time-independent simulation\cite{Groth2014_NJP} of the interchannel mixing with alternative shapes of the BS: (a) triangular potential dip and (b) rectangular potential dip, both $40$~nm long and with no extra smoothness. The BS is embedded in a device with only $L_1$, $L_2$ and $L_3$ leads (see Fig.~\ref{fig:kwantdevice}) to show that a coherent superposition of the first and the second channel is formed. In both cases, the beam is initialized in the first edge channel at $E=20.4$~meV and it is scattered to the second channel with about $(50\pm 4) \% $ probability by the (c) triangular potential dip and (d) rectangular potential dip. }\label{fig:bsalt}
\end{figure}

Finally, we present some support time-independent simulations performed with alternative shapes of the BS. We modeled a triangular and rectangular potential dip, whose profiles are reported in Fig.~\ref{fig:bsalt}(a) and Fig.~\ref{fig:bsalt}(b), respectively. In Fig.~\ref{fig:bsalt}(c)-(d), the two BSs are inserted in a simple device with the leads of injection and absorption, in order to show that they produce a coherent superposition of the first and the second channel. As in the previous case, the central energy of the WP can be chosen to obtain a $50 \%$ scattering probability between the two channels. We found that for both rectangular and triangular potential dips the scattering probability from the first to the second channel computed with Kwant software shows a small variation, around $5 \%$, for an energy dispersion of $0.2$~meV, which is comparable to the energy uncertainty usually obtained in experiments\cite{Marguerite2016_PRB}. 

\newpage

\begin{thebibliography}{49}%
\makeatletter
\providecommand \@ifxundefined [1]{%
 \@ifx{#1\undefined}
}%
\providecommand \@ifnum [1]{%
 \ifnum #1\expandafter \@firstoftwo
 \else \expandafter \@secondoftwo
 \fi
}%
\providecommand \@ifx [1]{%
 \ifx #1\expandafter \@firstoftwo
 \else \expandafter \@secondoftwo
 \fi
}%
\providecommand \natexlab [1]{#1}%
\providecommand \enquote  [1]{``#1''}%
\providecommand \bibnamefont  [1]{#1}%
\providecommand \bibfnamefont [1]{#1}%
\providecommand \citenamefont [1]{#1}%
\providecommand \href@noop [0]{\@secondoftwo}%
\providecommand \href [0]{\begingroup \@sanitize@url \@href}%
\providecommand \@href[1]{\@@startlink{#1}\@@href}%
\providecommand \@@href[1]{\endgroup#1\@@endlink}%
\providecommand \@sanitize@url [0]{\catcode `\\12\catcode `\$12\catcode
  `\&12\catcode `\#12\catcode `\^12\catcode `\_12\catcode `\%12\relax}%
\providecommand \@@startlink[1]{}%
\providecommand \@@endlink[0]{}%
\providecommand \url  [0]{\begingroup\@sanitize@url \@url }%
\providecommand \@url [1]{\endgroup\@href {#1}{\urlprefix }}%
\providecommand \urlprefix  [0]{URL }%
\providecommand \Eprint [0]{\href }%
\providecommand \doibase [0]{http://dx.doi.org/}%
\providecommand \selectlanguage [0]{\@gobble}%
\providecommand \bibinfo  [0]{\@secondoftwo}%
\providecommand \bibfield  [0]{\@secondoftwo}%
\providecommand \translation [1]{[#1]}%
\providecommand \BibitemOpen [0]{}%
\providecommand \bibitemStop [0]{}%
\providecommand \bibitemNoStop [0]{.\EOS\space}%
\providecommand \EOS [0]{\spacefactor3000\relax}%
\providecommand \BibitemShut  [1]{\csname bibitem#1\endcsname}%
\let\auto@bib@innerbib\@empty
\bibitem [{\citenamefont {Lanting}\ \emph {et~al.}(2014)\citenamefont
  {Lanting}, \citenamefont {Przybysz}, \citenamefont {Smirnov}, \citenamefont
  {Spedalieri}, \citenamefont {Amin}, \citenamefont {Berkley}, \citenamefont
  {Harris}, \citenamefont {Altomare}, \citenamefont {Boixo}, \citenamefont
  {Bunyk}, \citenamefont {Dickson}, \citenamefont {Enderud}, \citenamefont
  {Hilton}, \citenamefont {Hoskinson}, \citenamefont {Johnson}, \citenamefont
  {Ladizinsky}, \citenamefont {Ladizinsky}, \citenamefont {Neufeld},
  \citenamefont {Oh}, \citenamefont {Perminov}, \citenamefont {Rich},
  \citenamefont {Thom}, \citenamefont {Tolkacheva}, \citenamefont {Uchaikin},
  \citenamefont {Wilson},\ and\ \citenamefont {Rose}}]{Lanting2014_PRX}%
  \BibitemOpen
  \bibfield  {author} {\bibinfo {author} {\bibfnamefont {T.}~\bibnamefont
  {Lanting}}, \bibinfo {author} {\bibfnamefont {A.~J.}\ \bibnamefont
  {Przybysz}}, \bibinfo {author} {\bibfnamefont {A.~Y.}\ \bibnamefont
  {Smirnov}}, \bibinfo {author} {\bibfnamefont {F.~M.}\ \bibnamefont
  {Spedalieri}}, \bibinfo {author} {\bibfnamefont {M.~H.}\ \bibnamefont
  {Amin}}, \bibinfo {author} {\bibfnamefont {A.~J.}\ \bibnamefont {Berkley}},
  \bibinfo {author} {\bibfnamefont {R.}~\bibnamefont {Harris}}, \bibinfo
  {author} {\bibfnamefont {F.}~\bibnamefont {Altomare}}, \bibinfo {author}
  {\bibfnamefont {S.}~\bibnamefont {Boixo}}, \bibinfo {author} {\bibfnamefont
  {P.}~\bibnamefont {Bunyk}}, \bibinfo {author} {\bibfnamefont
  {N.}~\bibnamefont {Dickson}}, \bibinfo {author} {\bibfnamefont
  {C.}~\bibnamefont {Enderud}}, \bibinfo {author} {\bibfnamefont {J.~P.}\
  \bibnamefont {Hilton}}, \bibinfo {author} {\bibfnamefont {E.}~\bibnamefont
  {Hoskinson}}, \bibinfo {author} {\bibfnamefont {M.~W.}\ \bibnamefont
  {Johnson}}, \bibinfo {author} {\bibfnamefont {E.}~\bibnamefont {Ladizinsky}},
  \bibinfo {author} {\bibfnamefont {N.}~\bibnamefont {Ladizinsky}}, \bibinfo
  {author} {\bibfnamefont {R.}~\bibnamefont {Neufeld}}, \bibinfo {author}
  {\bibfnamefont {T.}~\bibnamefont {Oh}}, \bibinfo {author} {\bibfnamefont
  {I.}~\bibnamefont {Perminov}}, \bibinfo {author} {\bibfnamefont
  {C.}~\bibnamefont {Rich}}, \bibinfo {author} {\bibfnamefont {M.~C.}\
  \bibnamefont {Thom}}, \bibinfo {author} {\bibfnamefont {E.}~\bibnamefont
  {Tolkacheva}}, \bibinfo {author} {\bibfnamefont {S.}~\bibnamefont
  {Uchaikin}}, \bibinfo {author} {\bibfnamefont {A.~B.}\ \bibnamefont
  {Wilson}}, \ and\ \bibinfo {author} {\bibfnamefont {G.}~\bibnamefont
  {Rose}},\ }\href {\doibase 10.1103/PhysRevX.4.021041} {\bibfield  {journal}
  {\bibinfo  {journal} {Phys. Rev. X}\ }\textbf {\bibinfo {volume} {4}},\
  \bibinfo {pages} {021041} (\bibinfo {year} {2014})}\BibitemShut {NoStop}%
\bibitem [{\citenamefont {Debnath}\ \emph {et~al.}(2016)\citenamefont
  {Debnath}, \citenamefont {Linke}, \citenamefont {Figgatt}, \citenamefont
  {Landsman}, \citenamefont {Wright},\ and\ \citenamefont
  {Monroe}}]{Debnath2016_N}%
  \BibitemOpen
  \bibfield  {author} {\bibinfo {author} {\bibfnamefont {S.}~\bibnamefont
  {Debnath}}, \bibinfo {author} {\bibfnamefont {N.~M.}\ \bibnamefont {Linke}},
  \bibinfo {author} {\bibfnamefont {C.}~\bibnamefont {Figgatt}}, \bibinfo
  {author} {\bibfnamefont {K.~A.}\ \bibnamefont {Landsman}}, \bibinfo {author}
  {\bibfnamefont {K.}~\bibnamefont {Wright}}, \ and\ \bibinfo {author}
  {\bibfnamefont {C.}~\bibnamefont {Monroe}},\ }\href {\doibase
  http://dx.doi.org/10.1038/nature18648} {\bibfield  {journal} {\bibinfo
  {journal} {Nature}\ }\textbf {\bibinfo {volume} {536}},\ \bibinfo {pages}
  {63} (\bibinfo {year} {2016})}\BibitemShut {NoStop}%
\bibitem [{\citenamefont {Benenti}\ \emph {et~al.}(2004)\citenamefont
  {Benenti}, \citenamefont {Casati},\ and\ \citenamefont
  {Strini}}]{Benenti2004}%
  \BibitemOpen
  \bibfield  {author} {\bibinfo {author} {\bibfnamefont {G.}~\bibnamefont
  {Benenti}}, \bibinfo {author} {\bibfnamefont {G.}~\bibnamefont {Casati}}, \
  and\ \bibinfo {author} {\bibfnamefont {G.}~\bibnamefont {Strini}},\
  }\href@noop {} {\emph {\bibinfo {title} {Principles of Quantum Computation
  and Information. Volume 1: Basic Concepts}}}\ (\bibinfo  {publisher} {World
  Scientific},\ \bibinfo {address} {Singapore},\ \bibinfo {year}
  {2004})\BibitemShut {NoStop}%
\bibitem [{\citenamefont {Bertoni}\ and\ \citenamefont
  {Meyers}(2009)}]{Bertoni2009}%
  \BibitemOpen
  \bibfield  {author} {\bibinfo {author} {\bibfnamefont {A.}~\bibnamefont
  {Bertoni}}\ and\ \bibinfo {author} {\bibfnamefont {R.~A.}\ \bibnamefont
  {Meyers}},\ }\href@noop {} {\emph {\bibinfo {title} {Encyclopedia of
  Complexity and Systems Science}}}\ (\bibinfo  {publisher} {Springer},\
  \bibinfo {address} {Berlin},\ \bibinfo {year} {2009})\BibitemShut {NoStop}%
\bibitem [{\citenamefont {Yamamoto}\ \emph {et~al.}(2012)\citenamefont
  {Yamamoto}, \citenamefont {Takada}, \citenamefont {Bauerle}, \citenamefont
  {Watanabe}, \citenamefont {Wieck},\ and\ \citenamefont
  {Tarucha}}]{Yamamoto2012_NN}%
  \BibitemOpen
  \bibfield  {author} {\bibinfo {author} {\bibfnamefont {M.}~\bibnamefont
  {Yamamoto}}, \bibinfo {author} {\bibfnamefont {S.}~\bibnamefont {Takada}},
  \bibinfo {author} {\bibfnamefont {C.}~\bibnamefont {Bauerle}}, \bibinfo
  {author} {\bibfnamefont {K.}~\bibnamefont {Watanabe}}, \bibinfo {author}
  {\bibfnamefont {A.~D.}\ \bibnamefont {Wieck}}, \ and\ \bibinfo {author}
  {\bibfnamefont {S.}~\bibnamefont {Tarucha}},\ }\href {\doibase
  http://dx.doi.org/10.1038/nnano.2012.28} {\bibfield  {journal} {\bibinfo
  {journal} {Nat Nano}\ }\textbf {\bibinfo {volume} {7}},\ \bibinfo {pages}
  {247} (\bibinfo {year} {2012})}\BibitemShut {NoStop}%
\bibitem [{\citenamefont {Sarma}\ and\ \citenamefont
  {Pinczuk}(1997)}]{Sarma1997}%
  \BibitemOpen
  \bibfield  {author} {\bibinfo {author} {\bibfnamefont {S.~D.}\ \bibnamefont
  {Sarma}}\ and\ \bibinfo {author} {\bibfnamefont {A.}~\bibnamefont
  {Pinczuk}},\ }\href@noop {} {\emph {\bibinfo {title} {Perspectives in Quantum
  Hall Effects: Novel Quantum Liquids in Low-Dimensional Semiconductor
  Structures}}}\ (\bibinfo  {publisher} {Wiley, New York},\ \bibinfo {year}
  {1997})\BibitemShut {NoStop}%
\bibitem [{\citenamefont {Giovannetti}\ \emph {et~al.}(2008)\citenamefont
  {Giovannetti}, \citenamefont {Taddei}, \citenamefont {Frustaglia},\ and\
  \citenamefont {Fazio}}]{Giovannetti2008_PRB}%
  \BibitemOpen
  \bibfield  {author} {\bibinfo {author} {\bibfnamefont {V.}~\bibnamefont
  {Giovannetti}}, \bibinfo {author} {\bibfnamefont {F.}~\bibnamefont {Taddei}},
  \bibinfo {author} {\bibfnamefont {D.}~\bibnamefont {Frustaglia}}, \ and\
  \bibinfo {author} {\bibfnamefont {R.}~\bibnamefont {Fazio}},\ }\href
  {\doibase 10.1103/PhysRevB.77.155320} {\bibfield  {journal} {\bibinfo
  {journal} {Phys. Rev. B}\ }\textbf {\bibinfo {volume} {77}},\ \bibinfo
  {pages} {155320} (\bibinfo {year} {2008})}\BibitemShut {NoStop}%
\bibitem [{\citenamefont {Beggi}\ \emph {et~al.}(2015)\citenamefont {Beggi},
  \citenamefont {Bordone}, \citenamefont {Buscemi},\ and\ \citenamefont
  {Bertoni}}]{Beggi2015_JOPCM}%
  \BibitemOpen
  \bibfield  {author} {\bibinfo {author} {\bibfnamefont {A.}~\bibnamefont
  {Beggi}}, \bibinfo {author} {\bibfnamefont {P.}~\bibnamefont {Bordone}},
  \bibinfo {author} {\bibfnamefont {F.}~\bibnamefont {Buscemi}}, \ and\
  \bibinfo {author} {\bibfnamefont {A.}~\bibnamefont {Bertoni}},\ }\href@noop
  {} {\bibfield  {journal} {\bibinfo  {journal} {Journal of Physics: Condensed
  Matter}\ }\textbf {\bibinfo {volume} {27}},\ \bibinfo {pages} {475301}
  (\bibinfo {year} {2015})}\BibitemShut {NoStop}%
\bibitem [{\citenamefont {Venturelli}\ \emph {et~al.}(2011)\citenamefont
  {Venturelli}, \citenamefont {Giovannetti}, \citenamefont {Taddei},
  \citenamefont {Fazio}, \citenamefont {Feinberg}, \citenamefont {Usaj},\ and\
  \citenamefont {Balseiro}}]{Venturelli2011_PRB}%
  \BibitemOpen
  \bibfield  {author} {\bibinfo {author} {\bibfnamefont {D.}~\bibnamefont
  {Venturelli}}, \bibinfo {author} {\bibfnamefont {V.}~\bibnamefont
  {Giovannetti}}, \bibinfo {author} {\bibfnamefont {F.}~\bibnamefont {Taddei}},
  \bibinfo {author} {\bibfnamefont {R.}~\bibnamefont {Fazio}}, \bibinfo
  {author} {\bibfnamefont {D.}~\bibnamefont {Feinberg}}, \bibinfo {author}
  {\bibfnamefont {G.}~\bibnamefont {Usaj}}, \ and\ \bibinfo {author}
  {\bibfnamefont {C.~A.}\ \bibnamefont {Balseiro}},\ }\href {\doibase
  10.1103/PhysRevB.83.075315} {\bibfield  {journal} {\bibinfo  {journal} {Phys.
  Rev. B}\ }\textbf {\bibinfo {volume} {83}},\ \bibinfo {pages} {075315}
  (\bibinfo {year} {2011})}\BibitemShut {NoStop}%
\bibitem [{\citenamefont {Roulleau}\ \emph {et~al.}(2008)\citenamefont
  {Roulleau}, \citenamefont {Portier}, \citenamefont {Roche}, \citenamefont
  {Cavanna}, \citenamefont {Faini}, \citenamefont {Gennser},\ and\
  \citenamefont {Mailly}}]{Rolleau2008_PRL}%
  \BibitemOpen
  \bibfield  {author} {\bibinfo {author} {\bibfnamefont {P.}~\bibnamefont
  {Roulleau}}, \bibinfo {author} {\bibfnamefont {F.}~\bibnamefont {Portier}},
  \bibinfo {author} {\bibfnamefont {P.}~\bibnamefont {Roche}}, \bibinfo
  {author} {\bibfnamefont {A.}~\bibnamefont {Cavanna}}, \bibinfo {author}
  {\bibfnamefont {G.}~\bibnamefont {Faini}}, \bibinfo {author} {\bibfnamefont
  {U.}~\bibnamefont {Gennser}}, \ and\ \bibinfo {author} {\bibfnamefont
  {D.}~\bibnamefont {Mailly}},\ }\href {\doibase
  10.1103/PhysRevLett.100.126802} {\bibfield  {journal} {\bibinfo  {journal}
  {Phys. Rev. Lett.}\ }\textbf {\bibinfo {volume} {100}},\ \bibinfo {pages}
  {126802} (\bibinfo {year} {2008})}\BibitemShut {NoStop}%
\bibitem [{\citenamefont {Ji}\ \emph {et~al.}(2003)\citenamefont {Ji},
  \citenamefont {Chung}, \citenamefont {Sprinzak}, \citenamefont {Heiblum},
  \citenamefont {Mahalu},\ and\ \citenamefont {Shtrikman}}]{Ji2003_N}%
  \BibitemOpen
  \bibfield  {author} {\bibinfo {author} {\bibfnamefont {Y.}~\bibnamefont
  {Ji}}, \bibinfo {author} {\bibfnamefont {Y.}~\bibnamefont {Chung}}, \bibinfo
  {author} {\bibfnamefont {D.}~\bibnamefont {Sprinzak}}, \bibinfo {author}
  {\bibfnamefont {M.}~\bibnamefont {Heiblum}}, \bibinfo {author} {\bibfnamefont
  {D.}~\bibnamefont {Mahalu}}, \ and\ \bibinfo {author} {\bibfnamefont
  {H.}~\bibnamefont {Shtrikman}},\ }\href {\doibase
  http://dx.doi.org/doi:10.1038/nature01503} {\bibfield  {journal} {\bibinfo
  {journal} {Nature}\ }\textbf {\bibinfo {volume} {422}},\ \bibinfo {pages}
  {415} (\bibinfo {year} {2003})}\BibitemShut {NoStop}%
\bibitem [{\citenamefont {Deviatov}\ \emph {et~al.}(2011)\citenamefont
  {Deviatov}, \citenamefont {Ganczarczyk}, \citenamefont {Lorke}, \citenamefont
  {Biasiol},\ and\ \citenamefont {Sorba}}]{Deviatov2011_PRB}%
  \BibitemOpen
  \bibfield  {author} {\bibinfo {author} {\bibfnamefont {E.~V.}\ \bibnamefont
  {Deviatov}}, \bibinfo {author} {\bibfnamefont {A.}~\bibnamefont
  {Ganczarczyk}}, \bibinfo {author} {\bibfnamefont {A.}~\bibnamefont {Lorke}},
  \bibinfo {author} {\bibfnamefont {G.}~\bibnamefont {Biasiol}}, \ and\
  \bibinfo {author} {\bibfnamefont {L.}~\bibnamefont {Sorba}},\ }\href
  {\doibase 10.1103/PhysRevB.84.235313} {\bibfield  {journal} {\bibinfo
  {journal} {Phys. Rev. B}\ }\textbf {\bibinfo {volume} {84}},\ \bibinfo
  {pages} {235313} (\bibinfo {year} {2011})}\BibitemShut {NoStop}%
\bibitem [{\citenamefont {Deviatov}\ and\ \citenamefont
  {Lorke}(2008)}]{Deviatov2008_PRB}%
  \BibitemOpen
  \bibfield  {author} {\bibinfo {author} {\bibfnamefont {E.~V.}\ \bibnamefont
  {Deviatov}}\ and\ \bibinfo {author} {\bibfnamefont {A.}~\bibnamefont
  {Lorke}},\ }\href {\doibase 10.1103/PhysRevB.77.161302} {\bibfield  {journal}
  {\bibinfo  {journal} {Phys. Rev. B}\ }\textbf {\bibinfo {volume} {77}},\
  \bibinfo {pages} {161302} (\bibinfo {year} {2008})}\BibitemShut {NoStop}%
\bibitem [{\citenamefont {Deviatov}(2013)}]{Deviatov2013_LTP}%
  \BibitemOpen
  \bibfield  {author} {\bibinfo {author} {\bibfnamefont {E.~V.}\ \bibnamefont
  {Deviatov}},\ }\href {\doibase 10.1063/1.4775355} {\bibfield  {journal}
  {\bibinfo  {journal} {Low Temperature Physics}\ }\textbf {\bibinfo {volume}
  {39}},\ \bibinfo {pages} {7} (\bibinfo {year} {2013})}\BibitemShut {NoStop}%
\bibitem [{\citenamefont {Choi}\ \emph {et~al.}(2015)\citenamefont {Choi},
  \citenamefont {Sivan}, \citenamefont {Rosenblatt}, \citenamefont {Heiblum},
  \citenamefont {Umansky},\ and\ \citenamefont {Mahalu}}]{Choi2015_NC}%
  \BibitemOpen
  \bibfield  {author} {\bibinfo {author} {\bibfnamefont {H.}~\bibnamefont
  {Choi}}, \bibinfo {author} {\bibfnamefont {I.}~\bibnamefont {Sivan}},
  \bibinfo {author} {\bibfnamefont {A.}~\bibnamefont {Rosenblatt}}, \bibinfo
  {author} {\bibfnamefont {M.}~\bibnamefont {Heiblum}}, \bibinfo {author}
  {\bibfnamefont {V.}~\bibnamefont {Umansky}}, \ and\ \bibinfo {author}
  {\bibfnamefont {D.}~\bibnamefont {Mahalu}},\ }\href {\doibase
  10.1038/ncomms8435} {\bibfield  {journal} {\bibinfo  {journal} {Nature
  Communications}\ }\textbf {\bibinfo {volume} {6}},\ \bibinfo {pages} {7435}
  (\bibinfo {year} {2015})}\BibitemShut {NoStop}%
\bibitem [{\citenamefont {Bocquillon}\ \emph {et~al.}(2013)\citenamefont
  {Bocquillon}, \citenamefont {Freulon}, \citenamefont {Berroir}, \citenamefont
  {Degiovanni}, \citenamefont {Pla{\c c}ais}, \citenamefont {Cavanna},
  \citenamefont {Jin},\ and\ \citenamefont {F{\`e}ve}}]{Bocquillon2013_S}%
  \BibitemOpen
  \bibfield  {author} {\bibinfo {author} {\bibfnamefont {E.}~\bibnamefont
  {Bocquillon}}, \bibinfo {author} {\bibfnamefont {V.}~\bibnamefont {Freulon}},
  \bibinfo {author} {\bibfnamefont {J.-M.}\ \bibnamefont {Berroir}}, \bibinfo
  {author} {\bibfnamefont {P.}~\bibnamefont {Degiovanni}}, \bibinfo {author}
  {\bibfnamefont {B.}~\bibnamefont {Pla{\c c}ais}}, \bibinfo {author}
  {\bibfnamefont {A.}~\bibnamefont {Cavanna}}, \bibinfo {author} {\bibfnamefont
  {Y.}~\bibnamefont {Jin}}, \ and\ \bibinfo {author} {\bibfnamefont
  {G.}~\bibnamefont {F{\`e}ve}},\ }\href {\doibase 10.1126/science.1232572}
  {\bibfield  {journal} {\bibinfo  {journal} {Science}\ }\textbf {\bibinfo
  {volume} {339}},\ \bibinfo {pages} {1054} (\bibinfo {year}
  {2013})}\BibitemShut {NoStop}%
\bibitem [{\citenamefont {Oliver}\ \emph {et~al.}(1999)\citenamefont {Oliver},
  \citenamefont {Kim}, \citenamefont {Liu},\ and\ \citenamefont
  {Yamamoto}}]{Oliver1999_S}%
  \BibitemOpen
  \bibfield  {author} {\bibinfo {author} {\bibfnamefont {W.~D.}\ \bibnamefont
  {Oliver}}, \bibinfo {author} {\bibfnamefont {J.}~\bibnamefont {Kim}},
  \bibinfo {author} {\bibfnamefont {R.~C.}\ \bibnamefont {Liu}}, \ and\
  \bibinfo {author} {\bibfnamefont {Y.}~\bibnamefont {Yamamoto}},\ }\href
  {\doibase 10.1126/science.284.5412.299} {\bibfield  {journal} {\bibinfo
  {journal} {Science}\ }\textbf {\bibinfo {volume} {284}},\ \bibinfo {pages}
  {299} (\bibinfo {year} {1999})}\BibitemShut {NoStop}%
\bibitem [{\citenamefont {Marian}\ \emph {et~al.}(2015)\citenamefont {Marian},
  \citenamefont {Colomés},\ and\ \citenamefont {Oriols}}]{Marian2015_JOP}%
  \BibitemOpen
  \bibfield  {author} {\bibinfo {author} {\bibfnamefont {D.}~\bibnamefont
  {Marian}}, \bibinfo {author} {\bibfnamefont {E.}~\bibnamefont {Colomés}}, \
  and\ \bibinfo {author} {\bibfnamefont {X.}~\bibnamefont {Oriols}},\
  }\href@noop {} {\bibfield  {journal} {\bibinfo  {journal} {Journal of
  Physics: Condensed Matter}\ }\textbf {\bibinfo {volume} {27}},\ \bibinfo
  {pages} {245302} (\bibinfo {year} {2015})}\BibitemShut {NoStop}%
\bibitem [{\citenamefont {Marguerite}\ \emph {et~al.}(2016)\citenamefont
  {Marguerite}, \citenamefont {Cabart}, \citenamefont {Wahl}, \citenamefont
  {Roussel}, \citenamefont {Freulon}, \citenamefont {Ferraro}, \citenamefont
  {Grenier}, \citenamefont {Berroir}, \citenamefont
  {Pla\ifmmode~\mbox{\c{c}}\else \c{c}\fi{}ais}, \citenamefont {Jonckheere},
  \citenamefont {Rech}, \citenamefont {Martin}, \citenamefont {Degiovanni},
  \citenamefont {Cavanna}, \citenamefont {Jin},\ and\ \citenamefont
  {F\`eve}}]{Marguerite2016_PRB}%
  \BibitemOpen
  \bibfield  {author} {\bibinfo {author} {\bibfnamefont {A.}~\bibnamefont
  {Marguerite}}, \bibinfo {author} {\bibfnamefont {C.}~\bibnamefont {Cabart}},
  \bibinfo {author} {\bibfnamefont {C.}~\bibnamefont {Wahl}}, \bibinfo {author}
  {\bibfnamefont {B.}~\bibnamefont {Roussel}}, \bibinfo {author} {\bibfnamefont
  {V.}~\bibnamefont {Freulon}}, \bibinfo {author} {\bibfnamefont
  {D.}~\bibnamefont {Ferraro}}, \bibinfo {author} {\bibfnamefont
  {C.}~\bibnamefont {Grenier}}, \bibinfo {author} {\bibfnamefont {J.-M.}\
  \bibnamefont {Berroir}}, \bibinfo {author} {\bibfnamefont {B.}~\bibnamefont
  {Pla\ifmmode~\mbox{\c{c}}\else \c{c}\fi{}ais}}, \bibinfo {author}
  {\bibfnamefont {T.}~\bibnamefont {Jonckheere}}, \bibinfo {author}
  {\bibfnamefont {J.}~\bibnamefont {Rech}}, \bibinfo {author} {\bibfnamefont
  {T.}~\bibnamefont {Martin}}, \bibinfo {author} {\bibfnamefont
  {P.}~\bibnamefont {Degiovanni}}, \bibinfo {author} {\bibfnamefont
  {A.}~\bibnamefont {Cavanna}}, \bibinfo {author} {\bibfnamefont
  {Y.}~\bibnamefont {Jin}}, \ and\ \bibinfo {author} {\bibfnamefont
  {G.}~\bibnamefont {F\`eve}},\ }\href {\doibase 10.1103/PhysRevB.94.115311}
  {\bibfield  {journal} {\bibinfo  {journal} {Phys. Rev. B}\ }\textbf {\bibinfo
  {volume} {94}},\ \bibinfo {pages} {115311} (\bibinfo {year}
  {2016})}\BibitemShut {NoStop}%
\bibitem [{\citenamefont {Freulon}\ \emph {et~al.}(2015)\citenamefont
  {Freulon}, \citenamefont {Marguerite}, \citenamefont {Berroir}, \citenamefont
  {Plaçais}, \citenamefont {Cavanna}, \citenamefont {Jin},\ and\ \citenamefont
  {F{\`e}ve}}]{Marguerite2015_NatComm}%
  \BibitemOpen
  \bibfield  {author} {\bibinfo {author} {\bibfnamefont {V.}~\bibnamefont
  {Freulon}}, \bibinfo {author} {\bibfnamefont {A.}~\bibnamefont {Marguerite}},
  \bibinfo {author} {\bibfnamefont {J.-M.}\ \bibnamefont {Berroir}}, \bibinfo
  {author} {\bibfnamefont {B.}~\bibnamefont {Plaçais}}, \bibinfo {author}
  {\bibfnamefont {A.}~\bibnamefont {Cavanna}}, \bibinfo {author} {\bibfnamefont
  {Y.}~\bibnamefont {Jin}}, \ and\ \bibinfo {author} {\bibfnamefont
  {G.}~\bibnamefont {F{\`e}ve}},\ }\href {\doibase 10.1038/ncomms7854}
  {\bibfield  {journal} {\bibinfo  {journal} {Nature Communications}\ }\textbf
  {\bibinfo {volume} {6}},\ \bibinfo {pages} {6854} (\bibinfo {year}
  {2015})}\BibitemShut {NoStop}%
\bibitem [{\citenamefont {Neder}\ \emph
  {et~al.}(2007{\natexlab{a}})\citenamefont {Neder}, \citenamefont {Ofek},
  \citenamefont {Chung}, \citenamefont {Heiblum}, \citenamefont {Mahalu},\ and\
  \citenamefont {Umansky}}]{Neder2007_N}%
  \BibitemOpen
  \bibfield  {author} {\bibinfo {author} {\bibfnamefont {I.}~\bibnamefont
  {Neder}}, \bibinfo {author} {\bibfnamefont {N.}~\bibnamefont {Ofek}},
  \bibinfo {author} {\bibfnamefont {Y.}~\bibnamefont {Chung}}, \bibinfo
  {author} {\bibfnamefont {M.}~\bibnamefont {Heiblum}}, \bibinfo {author}
  {\bibfnamefont {D.}~\bibnamefont {Mahalu}}, \ and\ \bibinfo {author}
  {\bibfnamefont {V.}~\bibnamefont {Umansky}},\ }\href {\doibase
  10.1038/nature05955} {\bibfield  {journal} {\bibinfo  {journal} {Nature}\
  }\textbf {\bibinfo {volume} {448}},\ \bibinfo {pages} {333} (\bibinfo {year}
  {2007}{\natexlab{a}})}\BibitemShut {NoStop}%
\bibitem [{\citenamefont {Bocquillon}\ \emph {et~al.}(2012)\citenamefont
  {Bocquillon}, \citenamefont {Parmentier}, \citenamefont {Grenier},
  \citenamefont {Berroir}, \citenamefont {Degiovanni}, \citenamefont {Glattli},
  \citenamefont {Pla\ifmmode~\mbox{\c{c}}\else \c{c}\fi{}ais}, \citenamefont
  {Cavanna}, \citenamefont {Jin},\ and\ \citenamefont
  {F\`eve}}]{Bocquillon2012_PRL}%
  \BibitemOpen
  \bibfield  {author} {\bibinfo {author} {\bibfnamefont {E.}~\bibnamefont
  {Bocquillon}}, \bibinfo {author} {\bibfnamefont {F.~D.}\ \bibnamefont
  {Parmentier}}, \bibinfo {author} {\bibfnamefont {C.}~\bibnamefont {Grenier}},
  \bibinfo {author} {\bibfnamefont {J.-M.}\ \bibnamefont {Berroir}}, \bibinfo
  {author} {\bibfnamefont {P.}~\bibnamefont {Degiovanni}}, \bibinfo {author}
  {\bibfnamefont {D.~C.}\ \bibnamefont {Glattli}}, \bibinfo {author}
  {\bibfnamefont {B.}~\bibnamefont {Pla\ifmmode~\mbox{\c{c}}\else
  \c{c}\fi{}ais}}, \bibinfo {author} {\bibfnamefont {A.}~\bibnamefont
  {Cavanna}}, \bibinfo {author} {\bibfnamefont {Y.}~\bibnamefont {Jin}}, \ and\
  \bibinfo {author} {\bibfnamefont {G.}~\bibnamefont {F\`eve}},\ }\href
  {\doibase 10.1103/PhysRevLett.108.196803} {\bibfield  {journal} {\bibinfo
  {journal} {Phys. Rev. Lett.}\ }\textbf {\bibinfo {volume} {108}},\ \bibinfo
  {pages} {196803} (\bibinfo {year} {2012})}\BibitemShut {NoStop}%
\bibitem [{\citenamefont {Weisz}\ \emph {et~al.}(2014)\citenamefont {Weisz},
  \citenamefont {Choi}, \citenamefont {Sivan}, \citenamefont {Heiblum},
  \citenamefont {Gefen}, \citenamefont {Mahalu},\ and\ \citenamefont
  {Umansky}}]{Weisz2014_S}%
  \BibitemOpen
  \bibfield  {author} {\bibinfo {author} {\bibfnamefont {E.}~\bibnamefont
  {Weisz}}, \bibinfo {author} {\bibfnamefont {H.~K.}\ \bibnamefont {Choi}},
  \bibinfo {author} {\bibfnamefont {I.}~\bibnamefont {Sivan}}, \bibinfo
  {author} {\bibfnamefont {M.}~\bibnamefont {Heiblum}}, \bibinfo {author}
  {\bibfnamefont {Y.}~\bibnamefont {Gefen}}, \bibinfo {author} {\bibfnamefont
  {D.}~\bibnamefont {Mahalu}}, \ and\ \bibinfo {author} {\bibfnamefont
  {V.}~\bibnamefont {Umansky}},\ }\href {\doibase 10.1126/science.1248459}
  {\bibfield  {journal} {\bibinfo  {journal} {Science}\ }\textbf {\bibinfo
  {volume} {344}},\ \bibinfo {pages} {1363} (\bibinfo {year}
  {2014})}\BibitemShut {NoStop}%
\bibitem [{\citenamefont {Neder}\ \emph
  {et~al.}(2007{\natexlab{b}})\citenamefont {Neder}, \citenamefont {Heiblum},
  \citenamefont {Mahalu},\ and\ \citenamefont {Umansky}}]{Neder2007_PRL}%
  \BibitemOpen
  \bibfield  {author} {\bibinfo {author} {\bibfnamefont {I.}~\bibnamefont
  {Neder}}, \bibinfo {author} {\bibfnamefont {M.}~\bibnamefont {Heiblum}},
  \bibinfo {author} {\bibfnamefont {D.}~\bibnamefont {Mahalu}}, \ and\ \bibinfo
  {author} {\bibfnamefont {V.}~\bibnamefont {Umansky}},\ }\href {\doibase
  10.1103/PhysRevLett.98.036803} {\bibfield  {journal} {\bibinfo  {journal}
  {Phys. Rev. Lett.}\ }\textbf {\bibinfo {volume} {98}},\ \bibinfo {pages}
  {036803} (\bibinfo {year} {2007}{\natexlab{b}})}\BibitemShut {NoStop}%
\bibitem [{\citenamefont {Kreisbeck}\ \emph {et~al.}(2010)\citenamefont
  {Kreisbeck}, \citenamefont {Kramer}, \citenamefont {Buchholz}, \citenamefont
  {Fischer}, \citenamefont {Kunze}, \citenamefont {Reuter},\ and\ \citenamefont
  {Wieck}}]{Kreisbeck2010_PRB}%
  \BibitemOpen
  \bibfield  {author} {\bibinfo {author} {\bibfnamefont {C.}~\bibnamefont
  {Kreisbeck}}, \bibinfo {author} {\bibfnamefont {T.}~\bibnamefont {Kramer}},
  \bibinfo {author} {\bibfnamefont {S.~S.}\ \bibnamefont {Buchholz}}, \bibinfo
  {author} {\bibfnamefont {S.~F.}\ \bibnamefont {Fischer}}, \bibinfo {author}
  {\bibfnamefont {U.}~\bibnamefont {Kunze}}, \bibinfo {author} {\bibfnamefont
  {D.}~\bibnamefont {Reuter}}, \ and\ \bibinfo {author} {\bibfnamefont {A.~D.}\
  \bibnamefont {Wieck}},\ }\href {\doibase 10.1103/PhysRevB.82.165329}
  {\bibfield  {journal} {\bibinfo  {journal} {Phys. Rev. B}\ }\textbf {\bibinfo
  {volume} {82}},\ \bibinfo {pages} {165329} (\bibinfo {year}
  {2010})}\BibitemShut {NoStop}%
\bibitem [{\citenamefont {Palacios}\ and\ \citenamefont
  {Tejedor}(1993)}]{Palacios1993_PRB}%
  \BibitemOpen
  \bibfield  {author} {\bibinfo {author} {\bibfnamefont {J.~J.}\ \bibnamefont
  {Palacios}}\ and\ \bibinfo {author} {\bibfnamefont {C.}~\bibnamefont
  {Tejedor}},\ }\href@noop {} {\bibfield  {journal} {\bibinfo  {journal} {Phys.
  Rev. B}\ }\textbf {\bibinfo {volume} {48}},\ \bibinfo {pages} {5386}
  (\bibinfo {year} {1993})}\BibitemShut {NoStop}%
\bibitem [{\citenamefont {B.~Gaury}\ and\ \citenamefont
  {Waintal}(2015)}]{flyingqb_Waintal}%
  \BibitemOpen
  \bibfield  {author} {\bibinfo {author} {\bibfnamefont {J.~W.}\ \bibnamefont
  {B.~Gaury}}\ and\ \bibinfo {author} {\bibfnamefont {X.}~\bibnamefont
  {Waintal}},\ }\href@noop {} {\bibfield  {journal} {\bibinfo  {journal}
  {Nature Communications}\ }\textbf {\bibinfo {volume} {6}},\ \bibinfo {pages}
  {6524} (\bibinfo {year} {2015})}\BibitemShut {NoStop}%
\bibitem [{\citenamefont {Kramer}\ \emph {et~al.}(2010)\citenamefont {Kramer},
  \citenamefont {Kreisbeck},\ and\ \citenamefont {Krueckl}}]{Kramer2010_PS}%
  \BibitemOpen
  \bibfield  {author} {\bibinfo {author} {\bibfnamefont {T.}~\bibnamefont
  {Kramer}}, \bibinfo {author} {\bibfnamefont {C.}~\bibnamefont {Kreisbeck}}, \
  and\ \bibinfo {author} {\bibfnamefont {V.}~\bibnamefont {Krueckl}},\
  }\href@noop {} {\bibfield  {journal} {\bibinfo  {journal} {Physica Scripta}\
  }\textbf {\bibinfo {volume} {82}},\ \bibinfo {pages} {038101} (\bibinfo
  {year} {2010})}\BibitemShut {NoStop}%
\bibitem [{\citenamefont {Paradiso}\ \emph {et~al.}(2012)\citenamefont
  {Paradiso}, \citenamefont {Heun}, \citenamefont {Roddaro}, \citenamefont
  {Biasiol}, \citenamefont {Sorba}, \citenamefont {Venturelli}, \citenamefont
  {Taddei}, \citenamefont {Giovannetti},\ and\ \citenamefont
  {Beltram}}]{Paradiso2012_PRB}%
  \BibitemOpen
  \bibfield  {author} {\bibinfo {author} {\bibfnamefont {N.}~\bibnamefont
  {Paradiso}}, \bibinfo {author} {\bibfnamefont {S.}~\bibnamefont {Heun}},
  \bibinfo {author} {\bibfnamefont {S.}~\bibnamefont {Roddaro}}, \bibinfo
  {author} {\bibfnamefont {G.}~\bibnamefont {Biasiol}}, \bibinfo {author}
  {\bibfnamefont {L.}~\bibnamefont {Sorba}}, \bibinfo {author} {\bibfnamefont
  {D.}~\bibnamefont {Venturelli}}, \bibinfo {author} {\bibfnamefont
  {F.}~\bibnamefont {Taddei}}, \bibinfo {author} {\bibfnamefont
  {V.}~\bibnamefont {Giovannetti}}, \ and\ \bibinfo {author} {\bibfnamefont
  {F.}~\bibnamefont {Beltram}},\ }\href {\doibase 10.1103/PhysRevB.86.085326}
  {\bibfield  {journal} {\bibinfo  {journal} {Phys. Rev. B}\ }\textbf {\bibinfo
  {volume} {86}},\ \bibinfo {pages} {085326} (\bibinfo {year}
  {2012})}\BibitemShut {NoStop}%
\bibitem [{\citenamefont {Palacios}\ and\ \citenamefont
  {Tejedor}(1992)}]{Palacios1992_PRB}%
  \BibitemOpen
  \bibfield  {author} {\bibinfo {author} {\bibfnamefont {J.~J.}\ \bibnamefont
  {Palacios}}\ and\ \bibinfo {author} {\bibfnamefont {C.}~\bibnamefont
  {Tejedor}},\ }\href@noop {} {\bibfield  {journal} {\bibinfo  {journal} {Phys.
  Rev. B}\ }\textbf {\bibinfo {volume} {45}},\ \bibinfo {pages} {9059}
  (\bibinfo {year} {1992})}\BibitemShut {NoStop}%
\bibitem [{\citenamefont {Karmakar}\ \emph {et~al.}(2013)\citenamefont
  {Karmakar}, \citenamefont {Venturelli}, \citenamefont {Chirolli},
  \citenamefont {Taddei}, \citenamefont {Giovannetti}, \citenamefont {Fazio},
  \citenamefont {Roddaro}, \citenamefont {Biasiol}, \citenamefont {Sorba},
  \citenamefont {Pfeiffer}, \citenamefont {West}, \citenamefont {Pellegrini},\
  and\ \citenamefont {Beltram}}]{Karmakar2013_JOPCS}%
  \BibitemOpen
  \bibfield  {author} {\bibinfo {author} {\bibfnamefont {B.}~\bibnamefont
  {Karmakar}}, \bibinfo {author} {\bibfnamefont {D.}~\bibnamefont
  {Venturelli}}, \bibinfo {author} {\bibfnamefont {L.}~\bibnamefont
  {Chirolli}}, \bibinfo {author} {\bibfnamefont {F.}~\bibnamefont {Taddei}},
  \bibinfo {author} {\bibfnamefont {V.}~\bibnamefont {Giovannetti}}, \bibinfo
  {author} {\bibfnamefont {R.}~\bibnamefont {Fazio}}, \bibinfo {author}
  {\bibfnamefont {S.}~\bibnamefont {Roddaro}}, \bibinfo {author} {\bibfnamefont
  {G.}~\bibnamefont {Biasiol}}, \bibinfo {author} {\bibfnamefont
  {L.}~\bibnamefont {Sorba}}, \bibinfo {author} {\bibfnamefont {L.~N.}\
  \bibnamefont {Pfeiffer}}, \bibinfo {author} {\bibfnamefont {K.~W.}\
  \bibnamefont {West}}, \bibinfo {author} {\bibfnamefont {V.}~\bibnamefont
  {Pellegrini}}, \ and\ \bibinfo {author} {\bibfnamefont {F.}~\bibnamefont
  {Beltram}},\ }\href@noop {} {\bibfield  {journal} {\bibinfo  {journal}
  {Journal of Physics: Conference Series}\ }\textbf {\bibinfo {volume} {456}},\
  \bibinfo {pages} {012019} (\bibinfo {year} {2013})}\BibitemShut {NoStop}%
\bibitem [{\citenamefont {Karmakar}\ \emph {et~al.}(2015)\citenamefont
  {Karmakar}, \citenamefont {Venturelli}, \citenamefont {Chirolli},
  \citenamefont {Giovannetti}, \citenamefont {Fazio}, \citenamefont {Roddaro},
  \citenamefont {Pfeiffer}, \citenamefont {West}, \citenamefont {Taddei},\ and\
  \citenamefont {Pellegrini}}]{Karmakar2015_PRB}%
  \BibitemOpen
  \bibfield  {author} {\bibinfo {author} {\bibfnamefont {B.}~\bibnamefont
  {Karmakar}}, \bibinfo {author} {\bibfnamefont {D.}~\bibnamefont
  {Venturelli}}, \bibinfo {author} {\bibfnamefont {L.}~\bibnamefont
  {Chirolli}}, \bibinfo {author} {\bibfnamefont {V.}~\bibnamefont
  {Giovannetti}}, \bibinfo {author} {\bibfnamefont {R.}~\bibnamefont {Fazio}},
  \bibinfo {author} {\bibfnamefont {S.}~\bibnamefont {Roddaro}}, \bibinfo
  {author} {\bibfnamefont {L.~N.}\ \bibnamefont {Pfeiffer}}, \bibinfo {author}
  {\bibfnamefont {K.~W.}\ \bibnamefont {West}}, \bibinfo {author}
  {\bibfnamefont {F.}~\bibnamefont {Taddei}}, \ and\ \bibinfo {author}
  {\bibfnamefont {V.}~\bibnamefont {Pellegrini}},\ }\href {\doibase
  10.1103/PhysRevB.92.195303} {\bibfield  {journal} {\bibinfo  {journal} {Phys.
  Rev. B}\ }\textbf {\bibinfo {volume} {92}},\ \bibinfo {pages} {195303}
  (\bibinfo {year} {2015})}\BibitemShut {NoStop}%
\bibitem [{\citenamefont {Karmakar}\ \emph {et~al.}(2011)\citenamefont
  {Karmakar}, \citenamefont {Venturelli}, \citenamefont {Chirolli},
  \citenamefont {Taddei}, \citenamefont {Giovannetti}, \citenamefont {Fazio},
  \citenamefont {Roddaro}, \citenamefont {Biasiol}, \citenamefont {Sorba},
  \citenamefont {Pellegrini},\ and\ \citenamefont
  {Beltram}}]{Karmakar2011_PRL}%
  \BibitemOpen
  \bibfield  {author} {\bibinfo {author} {\bibfnamefont {B.}~\bibnamefont
  {Karmakar}}, \bibinfo {author} {\bibfnamefont {D.}~\bibnamefont
  {Venturelli}}, \bibinfo {author} {\bibfnamefont {L.}~\bibnamefont
  {Chirolli}}, \bibinfo {author} {\bibfnamefont {F.}~\bibnamefont {Taddei}},
  \bibinfo {author} {\bibfnamefont {V.}~\bibnamefont {Giovannetti}}, \bibinfo
  {author} {\bibfnamefont {R.}~\bibnamefont {Fazio}}, \bibinfo {author}
  {\bibfnamefont {S.}~\bibnamefont {Roddaro}}, \bibinfo {author} {\bibfnamefont
  {G.}~\bibnamefont {Biasiol}}, \bibinfo {author} {\bibfnamefont
  {L.}~\bibnamefont {Sorba}}, \bibinfo {author} {\bibfnamefont
  {V.}~\bibnamefont {Pellegrini}}, \ and\ \bibinfo {author} {\bibfnamefont
  {F.}~\bibnamefont {Beltram}},\ }\href {\doibase
  10.1103/PhysRevLett.107.236804} {\bibfield  {journal} {\bibinfo  {journal}
  {Phys. Rev. Lett.}\ }\textbf {\bibinfo {volume} {107}},\ \bibinfo {pages}
  {236804} (\bibinfo {year} {2011})}\BibitemShut {NoStop}%
\bibitem [{\citenamefont {F{\`e}ve}\ \emph {et~al.}(2007)\citenamefont
  {F{\`e}ve}, \citenamefont {Mahé}, \citenamefont {Berroir}, \citenamefont
  {Kontos}, \citenamefont {Plaçais}, \citenamefont {Glattli}, \citenamefont
  {Cavanna}, \citenamefont {Etienne},\ and\ \citenamefont {Jin}}]{Feve2007_S}%
  \BibitemOpen
  \bibfield  {author} {\bibinfo {author} {\bibfnamefont {G.}~\bibnamefont
  {F{\`e}ve}}, \bibinfo {author} {\bibfnamefont {A.}~\bibnamefont {Mahé}},
  \bibinfo {author} {\bibfnamefont {J.~M.}\ \bibnamefont {Berroir}}, \bibinfo
  {author} {\bibfnamefont {T.}~\bibnamefont {Kontos}}, \bibinfo {author}
  {\bibfnamefont {B.}~\bibnamefont {Plaçais}}, \bibinfo {author}
  {\bibfnamefont {D.~C.}\ \bibnamefont {Glattli}}, \bibinfo {author}
  {\bibfnamefont {A.}~\bibnamefont {Cavanna}}, \bibinfo {author} {\bibfnamefont
  {B.}~\bibnamefont {Etienne}}, \ and\ \bibinfo {author} {\bibfnamefont
  {Y.}~\bibnamefont {Jin}},\ }\href@noop {} {\bibfield  {journal} {\bibinfo
  {journal} {Science}\ }\textbf {\bibinfo {volume} {316}},\ \bibinfo {pages}
  {5828} (\bibinfo {year} {2007})}\BibitemShut {NoStop}%
\bibitem [{\citenamefont {Keeling}\ \emph {et~al.}(2008)\citenamefont
  {Keeling}, \citenamefont {Shytov},\ and\ \citenamefont
  {Levitov}}]{Keeling2008_PRL}%
  \BibitemOpen
  \bibfield  {author} {\bibinfo {author} {\bibfnamefont {J.}~\bibnamefont
  {Keeling}}, \bibinfo {author} {\bibfnamefont {A.~V.}\ \bibnamefont {Shytov}},
  \ and\ \bibinfo {author} {\bibfnamefont {L.~S.}\ \bibnamefont {Levitov}},\
  }\href {\doibase 10.1103/PhysRevLett.101.196404} {\bibfield  {journal}
  {\bibinfo  {journal} {Phys. Rev. Lett.}\ }\textbf {\bibinfo {volume} {101}},\
  \bibinfo {pages} {196404} (\bibinfo {year} {2008})}\BibitemShut {NoStop}%
\bibitem [{\citenamefont {Kataoka}\ \emph {et~al.}(2016)\citenamefont
  {Kataoka}, \citenamefont {Johnson}, \citenamefont {Emary}, \citenamefont
  {See}, \citenamefont {Griffiths}, \citenamefont {Jones}, \citenamefont
  {Farrer}, \citenamefont {Ritchie}, \citenamefont {Pepper},\ and\
  \citenamefont {Janssen}}]{Kataoka2016_PRL}%
  \BibitemOpen
  \bibfield  {author} {\bibinfo {author} {\bibfnamefont {M.}~\bibnamefont
  {Kataoka}}, \bibinfo {author} {\bibfnamefont {N.}~\bibnamefont {Johnson}},
  \bibinfo {author} {\bibfnamefont {C.}~\bibnamefont {Emary}}, \bibinfo
  {author} {\bibfnamefont {P.}~\bibnamefont {See}}, \bibinfo {author}
  {\bibfnamefont {J.~P.}\ \bibnamefont {Griffiths}}, \bibinfo {author}
  {\bibfnamefont {G.~A.~C.}\ \bibnamefont {Jones}}, \bibinfo {author}
  {\bibfnamefont {I.}~\bibnamefont {Farrer}}, \bibinfo {author} {\bibfnamefont
  {D.~A.}\ \bibnamefont {Ritchie}}, \bibinfo {author} {\bibfnamefont
  {M.}~\bibnamefont {Pepper}}, \ and\ \bibinfo {author} {\bibfnamefont {T.~J.
  B.~M.}\ \bibnamefont {Janssen}},\ }\href {\doibase
  10.1103/PhysRevLett.116.126803} {\bibfield  {journal} {\bibinfo  {journal}
  {Phys. Rev. Lett.}\ }\textbf {\bibinfo {volume} {116}},\ \bibinfo {pages}
  {126803} (\bibinfo {year} {2016})}\BibitemShut {NoStop}%
\bibitem [{\citenamefont {Emary}\ \emph {et~al.}(2016)\citenamefont {Emary},
  \citenamefont {Dyson}, \citenamefont {Ryu}, \citenamefont {Sim},\ and\
  \citenamefont {Kataoka}}]{Emary2016_PRB}%
  \BibitemOpen
  \bibfield  {author} {\bibinfo {author} {\bibfnamefont {C.}~\bibnamefont
  {Emary}}, \bibinfo {author} {\bibfnamefont {A.}~\bibnamefont {Dyson}},
  \bibinfo {author} {\bibfnamefont {S.}~\bibnamefont {Ryu}}, \bibinfo {author}
  {\bibfnamefont {H.-S.}\ \bibnamefont {Sim}}, \ and\ \bibinfo {author}
  {\bibfnamefont {M.}~\bibnamefont {Kataoka}},\ }\href {\doibase
  10.1103/PhysRevB.93.035436} {\bibfield  {journal} {\bibinfo  {journal} {Phys.
  Rev. B}\ }\textbf {\bibinfo {volume} {93}},\ \bibinfo {pages} {035436}
  (\bibinfo {year} {2016})}\BibitemShut {NoStop}%
\bibitem [{\citenamefont {Grasselli}\ \emph {et~al.}(2016)\citenamefont
  {Grasselli}, \citenamefont {Bertoni},\ and\ \citenamefont
  {Goldoni}}]{Grasselli2016_PRB}%
  \BibitemOpen
  \bibfield  {author} {\bibinfo {author} {\bibfnamefont {F.}~\bibnamefont
  {Grasselli}}, \bibinfo {author} {\bibfnamefont {A.}~\bibnamefont {Bertoni}},
  \ and\ \bibinfo {author} {\bibfnamefont {G.}~\bibnamefont {Goldoni}},\ }\href
  {\doibase 10.1103/PhysRevB.93.195310} {\bibfield  {journal} {\bibinfo
  {journal} {Phys. Rev. B}\ }\textbf {\bibinfo {volume} {93}},\ \bibinfo
  {pages} {195310} (\bibinfo {year} {2016})}\BibitemShut {NoStop}%
\bibitem [{\citenamefont {Groth}\ \emph {et~al.}(2014)\citenamefont {Groth},
  \citenamefont {Wimmer}, \citenamefont {Akhmerov},\ and\ \citenamefont
  {Waintal}}]{Groth2014_NJP}%
  \BibitemOpen
  \bibfield  {author} {\bibinfo {author} {\bibfnamefont {C.~W.}\ \bibnamefont
  {Groth}}, \bibinfo {author} {\bibfnamefont {M.}~\bibnamefont {Wimmer}},
  \bibinfo {author} {\bibfnamefont {A.~R.}\ \bibnamefont {Akhmerov}}, \ and\
  \bibinfo {author} {\bibfnamefont {X.}~\bibnamefont {Waintal}},\ }\href@noop
  {} {\bibfield  {journal} {\bibinfo  {journal} {New Journal of Physics}\
  }\textbf {\bibinfo {volume} {16}},\ \bibinfo {pages} {063065} (\bibinfo
  {year} {2014})}\BibitemShut {NoStop}%
\bibitem [{\citenamefont {Bird}\ \emph {et~al.}(1994)\citenamefont {Bird},
  \citenamefont {Ishibashi}, \citenamefont {Stopa}, \citenamefont {Aoyagi},\
  and\ \citenamefont {Sugano}}]{Bird1994_PRB}%
  \BibitemOpen
  \bibfield  {author} {\bibinfo {author} {\bibfnamefont {J.~P.}\ \bibnamefont
  {Bird}}, \bibinfo {author} {\bibfnamefont {K.}~\bibnamefont {Ishibashi}},
  \bibinfo {author} {\bibfnamefont {M.}~\bibnamefont {Stopa}}, \bibinfo
  {author} {\bibfnamefont {Y.}~\bibnamefont {Aoyagi}}, \ and\ \bibinfo {author}
  {\bibfnamefont {T.}~\bibnamefont {Sugano}},\ }\href {\doibase
  10.1103/PhysRevB.50.14983} {\bibfield  {journal} {\bibinfo  {journal} {Phys.
  Rev. B}\ }\textbf {\bibinfo {volume} {50}},\ \bibinfo {pages} {14983}
  (\bibinfo {year} {1994})}\BibitemShut {NoStop}%
\bibitem [{\citenamefont {Keeling}\ \emph {et~al.}(2006)\citenamefont
  {Keeling}, \citenamefont {Klich},\ and\ \citenamefont
  {Levitov}}]{Keeling2006_PRL}%
  \BibitemOpen
  \bibfield  {author} {\bibinfo {author} {\bibfnamefont {J.}~\bibnamefont
  {Keeling}}, \bibinfo {author} {\bibfnamefont {I.}~\bibnamefont {Klich}}, \
  and\ \bibinfo {author} {\bibfnamefont {L.~S.}\ \bibnamefont {Levitov}},\
  }\href {\doibase 10.1103/PhysRevLett.97.116403} {\bibfield  {journal}
  {\bibinfo  {journal} {Phys. Rev. Lett.}\ }\textbf {\bibinfo {volume} {97}},\
  \bibinfo {pages} {116403} (\bibinfo {year} {2006})}\BibitemShut {NoStop}%
\bibitem [{\citenamefont {Dubois}\ \emph {et~al.}(2013)\citenamefont {Dubois},
  \citenamefont {Jullien}, \citenamefont {Portier}, \citenamefont {Roche},
  \citenamefont {Cavanna}, \citenamefont {Jin}, \citenamefont {Wegscheider},
  \citenamefont {Roulleau},\ and\ \citenamefont {Glattli}}]{Dubois2013_N}%
  \BibitemOpen
  \bibfield  {author} {\bibinfo {author} {\bibfnamefont {J.}~\bibnamefont
  {Dubois}}, \bibinfo {author} {\bibfnamefont {T.}~\bibnamefont {Jullien}},
  \bibinfo {author} {\bibfnamefont {F.}~\bibnamefont {Portier}}, \bibinfo
  {author} {\bibfnamefont {P.}~\bibnamefont {Roche}}, \bibinfo {author}
  {\bibfnamefont {A.}~\bibnamefont {Cavanna}}, \bibinfo {author} {\bibfnamefont
  {Y.}~\bibnamefont {Jin}}, \bibinfo {author} {\bibfnamefont {W.}~\bibnamefont
  {Wegscheider}}, \bibinfo {author} {\bibfnamefont {P.}~\bibnamefont
  {Roulleau}}, \ and\ \bibinfo {author} {\bibfnamefont {D.~C.}\ \bibnamefont
  {Glattli}},\ }\href {\doibase 10.1038/nature12713} {\bibfield  {journal}
  {\bibinfo  {journal} {Nature}\ }\textbf {\bibinfo {volume} {502}},\ \bibinfo
  {pages} {659} (\bibinfo {year} {2013})}\BibitemShut {NoStop}%
\bibitem [{\citenamefont {Ryu}\ \emph {et~al.}(2016)\citenamefont {Ryu},
  \citenamefont {Kataoka},\ and\ \citenamefont {Sim}}]{Ryu2016_PRL}%
  \BibitemOpen
  \bibfield  {author} {\bibinfo {author} {\bibfnamefont {S.}~\bibnamefont
  {Ryu}}, \bibinfo {author} {\bibfnamefont {M.}~\bibnamefont {Kataoka}}, \ and\
  \bibinfo {author} {\bibfnamefont {H.-S.}\ \bibnamefont {Sim}},\ }\href
  {\doibase 10.1103/PhysRevLett.117.146802} {\bibfield  {journal} {\bibinfo
  {journal} {Phys. Rev. Lett.}\ }\textbf {\bibinfo {volume} {117}},\ \bibinfo
  {pages} {146802} (\bibinfo {year} {2016})}\BibitemShut {NoStop}%
\bibitem [{\citenamefont {Haack}\ \emph {et~al.}(2011)\citenamefont {Haack},
  \citenamefont {Moskalets}, \citenamefont {Splettstoesser},\ and\
  \citenamefont {B\"uttiker}}]{Haack2011_PRB}%
  \BibitemOpen
  \bibfield  {author} {\bibinfo {author} {\bibfnamefont {G.}~\bibnamefont
  {Haack}}, \bibinfo {author} {\bibfnamefont {M.}~\bibnamefont {Moskalets}},
  \bibinfo {author} {\bibfnamefont {J.}~\bibnamefont {Splettstoesser}}, \ and\
  \bibinfo {author} {\bibfnamefont {M.}~\bibnamefont {B\"uttiker}},\ }\href
  {\doibase 10.1103/PhysRevB.84.081303} {\bibfield  {journal} {\bibinfo
  {journal} {Phys. Rev. B}\ }\textbf {\bibinfo {volume} {84}},\ \bibinfo
  {pages} {081303} (\bibinfo {year} {2011})}\BibitemShut {NoStop}%
\bibitem [{\citenamefont {Neder}\ \emph {et~al.}(2006)\citenamefont {Neder},
  \citenamefont {Heiblum}, \citenamefont {Levinson}, \citenamefont {Mahalu},\
  and\ \citenamefont {Umansky}}]{Neder2006_PRL}%
  \BibitemOpen
  \bibfield  {author} {\bibinfo {author} {\bibfnamefont {I.}~\bibnamefont
  {Neder}}, \bibinfo {author} {\bibfnamefont {M.}~\bibnamefont {Heiblum}},
  \bibinfo {author} {\bibfnamefont {Y.}~\bibnamefont {Levinson}}, \bibinfo
  {author} {\bibfnamefont {D.}~\bibnamefont {Mahalu}}, \ and\ \bibinfo {author}
  {\bibfnamefont {V.}~\bibnamefont {Umansky}},\ }\href {\doibase
  10.1103/PhysRevLett.96.016804} {\bibfield  {journal} {\bibinfo  {journal}
  {Phys. Rev. Lett.}\ }\textbf {\bibinfo {volume} {96}},\ \bibinfo {pages}
  {016804} (\bibinfo {year} {2006})}\BibitemShut {NoStop}%
\bibitem [{\citenamefont {Huynh}\ \emph {et~al.}(2012)\citenamefont {Huynh},
  \citenamefont {Portier}, \citenamefont {le~Sueur}, \citenamefont {Faini},
  \citenamefont {Gennser}, \citenamefont {Mailly}, \citenamefont {Pierre},
  \citenamefont {Wegscheider},\ and\ \citenamefont {Roche}}]{Huyn2012_PRL}%
  \BibitemOpen
  \bibfield  {author} {\bibinfo {author} {\bibfnamefont {P.-A.}\ \bibnamefont
  {Huynh}}, \bibinfo {author} {\bibfnamefont {F.}~\bibnamefont {Portier}},
  \bibinfo {author} {\bibfnamefont {H.}~\bibnamefont {le~Sueur}}, \bibinfo
  {author} {\bibfnamefont {G.}~\bibnamefont {Faini}}, \bibinfo {author}
  {\bibfnamefont {U.}~\bibnamefont {Gennser}}, \bibinfo {author} {\bibfnamefont
  {D.}~\bibnamefont {Mailly}}, \bibinfo {author} {\bibfnamefont
  {F.}~\bibnamefont {Pierre}}, \bibinfo {author} {\bibfnamefont
  {W.}~\bibnamefont {Wegscheider}}, \ and\ \bibinfo {author} {\bibfnamefont
  {P.}~\bibnamefont {Roche}},\ }\href {\doibase 10.1103/PhysRevLett.108.256802}
  {\bibfield  {journal} {\bibinfo  {journal} {Phys. Rev. Lett.}\ }\textbf
  {\bibinfo {volume} {108}},\ \bibinfo {pages} {256802} (\bibinfo {year}
  {2012})}\BibitemShut {NoStop}%
\bibitem [{\citenamefont {Helzel}\ \emph {et~al.}(2015)\citenamefont {Helzel},
  \citenamefont {Litvin}, \citenamefont {Levkivskyi}, \citenamefont
  {Sukhorukov}, \citenamefont {Wegscheider},\ and\ \citenamefont
  {Strunk}}]{Helzel2015_PRB}%
  \BibitemOpen
  \bibfield  {author} {\bibinfo {author} {\bibfnamefont {A.}~\bibnamefont
  {Helzel}}, \bibinfo {author} {\bibfnamefont {L.~V.}\ \bibnamefont {Litvin}},
  \bibinfo {author} {\bibfnamefont {I.~P.}\ \bibnamefont {Levkivskyi}},
  \bibinfo {author} {\bibfnamefont {E.~V.}\ \bibnamefont {Sukhorukov}},
  \bibinfo {author} {\bibfnamefont {W.}~\bibnamefont {Wegscheider}}, \ and\
  \bibinfo {author} {\bibfnamefont {C.}~\bibnamefont {Strunk}},\ }\href
  {\doibase 10.1103/PhysRevB.91.245419} {\bibfield  {journal} {\bibinfo
  {journal} {Phys. Rev. B}\ }\textbf {\bibinfo {volume} {91}},\ \bibinfo
  {pages} {245419} (\bibinfo {year} {2015})}\BibitemShut {NoStop}%
\bibitem [{\citenamefont {Chirolli}\ \emph {et~al.}(2013)\citenamefont
  {Chirolli}, \citenamefont {Taddei}, \citenamefont {Fazio},\ and\
  \citenamefont {Giovannetti}}]{Chirolli2013_PRL}%
  \BibitemOpen
  \bibfield  {author} {\bibinfo {author} {\bibfnamefont {L.}~\bibnamefont
  {Chirolli}}, \bibinfo {author} {\bibfnamefont {F.}~\bibnamefont {Taddei}},
  \bibinfo {author} {\bibfnamefont {R.}~\bibnamefont {Fazio}}, \ and\ \bibinfo
  {author} {\bibfnamefont {V.}~\bibnamefont {Giovannetti}},\ }\href {\doibase
  10.1103/PhysRevLett.111.036801} {\bibfield  {journal} {\bibinfo  {journal}
  {Phys. Rev. Lett.}\ }\textbf {\bibinfo {volume} {111}},\ \bibinfo {pages}
  {036801} (\bibinfo {year} {2013})}\BibitemShut {NoStop}%
\bibitem [{\citenamefont {Ferraro}\ \emph {et~al.}(2014)\citenamefont
  {Ferraro}, \citenamefont {Roussel}, \citenamefont {Cabart}, \citenamefont
  {Thibierge}, \citenamefont {F\`eve}, \citenamefont {Grenier},\ and\
  \citenamefont {Degiovanni}}]{Ferraro2014_PRL}%
  \BibitemOpen
  \bibfield  {author} {\bibinfo {author} {\bibfnamefont {D.}~\bibnamefont
  {Ferraro}}, \bibinfo {author} {\bibfnamefont {B.}~\bibnamefont {Roussel}},
  \bibinfo {author} {\bibfnamefont {C.}~\bibnamefont {Cabart}}, \bibinfo
  {author} {\bibfnamefont {E.}~\bibnamefont {Thibierge}}, \bibinfo {author}
  {\bibfnamefont {G.}~\bibnamefont {F\`eve}}, \bibinfo {author} {\bibfnamefont
  {C.}~\bibnamefont {Grenier}}, \ and\ \bibinfo {author} {\bibfnamefont
  {P.}~\bibnamefont {Degiovanni}},\ }\href {\doibase
  10.1103/PhysRevLett.113.166403} {\bibfield  {journal} {\bibinfo  {journal}
  {Phys. Rev. Lett.}\ }\textbf {\bibinfo {volume} {113}},\ \bibinfo {pages}
  {166403} (\bibinfo {year} {2014})}\BibitemShut {NoStop}%
\end{thebibliography}
\bibliographystyle{apsrev4-1}
\end{document}